\documentclass[twocolumn]{aastex62}
\usepackage{gensymb} 
\newcommand{\sdssj}{SDSS~J141324+530527}
\newcommand{\oiii}{[O~III]$\lambda 5007$}
\newcommand{\hb}{H$\beta$}
\newcommand{\ha}{H$\alpha$}
\newcommand{\hg}{H$\gamma$}
\newcommand{\mgii}{MgII}
\newcommand{\hei}{HeI}
\newcommand{\heii}{HeII}
\newcommand{\ergs}{\ensuremath{{\rm erg\,s}}^{-1}}

\graphicspath{{./}{figures/}}

\shorttitle{A spectrally resolved hypervariable quasar}
\shortauthors{Dexter et al.}

\begin{document}

\title{The Sloan Digital Sky Survey Reverberation Mapping Project:\\Accretion and Broad Emission Line Physics from a Hypervariable Quasar}

\email{jason.dexter@colorado.edu}

\author[0000-0003-3903-0373]{Jason Dexter}
\affil{Max-Planck-Institut f\"{u}r extraterrestrische Physik, Giessenbachstr. 1, 85748 Garching, Germany}
\affil{JILA and Department of Astrophysical and Planetary Sciences, University of Colorado, Boulder, CO 80309, USA}

\author{Shuo Xin}
\affil{Max-Planck-Institut f\"{u}r extraterrestrische Physik, Giessenbachstr. 1, 85748 Garching, Germany}
\affil{School of Physics Sciences and Engineering,
Tongji University, Shanghai 200092, China}

\author[0000-0003-1659-7035]{Yue Shen}
\affil{Department of Astronomy, University of Illinois at Urbana-Champaign, Urbana, IL 61801, USA}
\affil{National Center for Supercomputing Applications, University of Illinois at Urbana-Champaign, Urbana, IL 61801, USA}

\author[0000-0001-9920-6057]{C. J. Grier}
\affil{Department of Astronomy and Astrophysics, The Pennsylvania State University, 525 Davey Laboratory, University Park, PA 16802, USA}
\affil{Institute for Gravitation and the Cosmos, The Pennsylvania State University, University Park, PA 16802, USA}
\affil{Steward Observatory, The University of Arizona, 933 North Cherry Avenue, Tucson, AZ 85721, USA} 

\author{Teng Liu}
\affil{Max-Planck-Institut f\"{u}r extraterrestrische Physik, Giessenbachstr. 1, 85748 Garching, Germany}

\author{Suvi Gezari}
\affil{Department of Astronomy, University of Maryland, College Park, MD 20742, USA}
\affil{Joint Space-Science Institute, University of Maryland, College Park, MD 20742, USA}

\author[0000-0002-3461-5228]{Ian D. McGreer}
\affil{Steward Observatory, The University of Arizona, 933 North Cherry Avenue, Tucson, AZ 85721, USA}

\author[0000-0002-0167-2453]{W. N. Brandt}
\affil{Department of Astronomy and Astrophysics, The Pennsylvania State University, 525 Davey Laboratory, University Park, PA 16802, USA}

\affil{Department of Physics, 104 Davey Laboratory, The Pennsylvania State University, University Park, PA 16802, USA}

\affil{Institute for Gravitation and the Cosmos, The Pennsylvania State University, University Park, PA 16802, USA}

\author{P. B. Hall}
\affil{Department of Physics and Astronomy, York University, Toronto, ON M3J 1P3, Canada}

\author[0000-0003-1728-0304]{Keith Horne}
\affil{SUPA Physics and Astronomy, University of St Andrews, Fife, KY16 9SS, Scotland, UK}

\author{Torben Simm}
\affil{Max-Planck-Institut f\"{u}r extraterrestrische Physik, Giessenbachstr. 1, 85748 Garching, Germany}

\author{Andrea Merloni}
\affil{Max-Planck-Institut f\"{u}r extraterrestrische Physik, Giessenbachstr. 1, 85748 Garching, Germany}

\author[0000-0002-8179-9445]{Paul J. Green}
\affil{Harvard Smithsonian Center for Astrophysics, 60 Garden Street, Cambridge, MA 02138, USA}

\author{M. Vivek}
\affil{Department of Astronomy and Astrophysics, The Pennsylvania State University, 525 Davey Laboratory, University Park, PA 16802, USA}

\author[0000-0002-1410-0470]{Jonathan R. Trump}
\affil{University of Connecticut, Department of Physics, 2152 Hillside Road, Unit 3046, Storrs, CT 06269, USA}

\author[0000-0002-0957-7151]{Yasaman Homayouni}
\affil{University of Connecticut, Department of Physics, 2152 Hillside Road, Unit 3046, Storrs, CT 06269, USA}

\author{B. M. Peterson}
\affil{Department of Astronomy, The Ohio State University, 140 W 18th Ave, Columbus, OH 43210, USA}
\affil{Center for Cosmology and AstroParticle Physics, The Ohio State University, 191 West Woodruff Ave., Columbus, OH 43210, USA }
\affil{Space Telescope Science Institute, 3700 San Martin Drive, Baltimore, MD 21218, USA}

\author{Donald P. Schneider}
\affil{Department of Astronomy and Astrophysics, The Pennsylvania State University, 525 Davey Laboratory, University Park, PA 16802, USA}
\affil{Institute for Gravitation and the Cosmos, The Pennsylvania
  State University, University Park, PA 16802, USA}

\author{K. Kinemuchi}
\affil{Apache Point Observatory and New Mexico State University, P.O. Box 59, Sunspot, NM, 88349-0059, USA}

\author{Kaike Pan}
\affil{Apache Point Observatory and New Mexico State University, P.O. Box 59, Sunspot, NM, 88349-0059, USA}

\author{Dmitry Bizyaev}
\affil{Apache Point Observatory and New Mexico State University, P.O. Box 59, Sunspot, NM, 88349-0059, USA}
\affil{Sternberg Astronomical Institute, Moscow State University, Moscow}




\begin{abstract}
We analyze extensive spectroscopic and photometric data of the
hypervariable quasar \sdssj\, (RMID 017) at $z=0.456$, an optical
``changing look" quasar from the Sloan Digital Sky Survey
Reverberation Mapping project that increased in optical luminosity by
a factor $\simeq 10$ between 2014 and 2017. The observed broad emission lines
all respond in luminosity and width to the changing optical continuum, as expected for
photoionization in a stratified, virialized broad emission line
region. The luminosity changes therefore result from
intrinsic changes in accretion power rather than variable obscuration. The variability is continuous and apparently stochastic, disfavoring an origin as a discrete event such as a tidal disruption flare or microlensing event. It is coordinated on day timescales with blue leading red, consistent with reprocessing powering the entire optical SED. We show that this process cannot work in a standard thin disk geometry on energetic grounds, and would instead require a large covering factor reprocessor. Disk instability models could potentially also explain the data, provided that the instability sets in near the inner radius of a geometrically thick accretion disk.
\end{abstract}



\section{Introduction} \label{sec:intro}

The continuum optical/UV emission of quasars has long been known to
vary, with typical rms fluctuations of  $\simeq 10-20\%$ correlated on
timescales of months to years
\citep[e.g.,][]{kelly2009,macleod2010,kozlowski2010}. This variability
is often correlated over a wide wavelength range
\citep[e.g.,][]{clavel1991,peterson1991}, which requires a propagation
speed $\gtrsim 0.1c$ \citep{krolik1991} if the emission region size scales with characteristic emission wavelength $\lambda$ 
as $R \propto \lambda^{4/3}$ predicted by standard thin accretion disk theory \citep{shakura1973}. This timescale is much shorter than expected for the propagation of accretion rate fluctuations through the disk
\citep{pringle1981} and those variations are usually associated with reprocessing of emission from near the black hole. Intensive monitoring campaigns have shown that the X-ray relationship is more complex than for the UV \citep{edelson2019}. When either the UV or X-rays are correlated with the response at longer optical wavelengths, the measured lag in a reprocessing model measures the wavelength-dependent emission region size \citep[e.g.,][]{mchardy2014,fausnaugh2016}.

A small fraction of quasars exhibit large optical luminosity variations by factors of $\gtrsim 2$
\citep[``hypervariable,"][]{rumbaugh2018}. Rare, optical ``changing look" AGN show appearance or disappearance of one or more broad emission lines. Those changes are associated with factor of $\sim 10$ variations in optical continuum luminosity on month to year timescales. Long known in nearby Seyferts
\citep{tohlineosterbrock1976,cohen1986,clavel1989,storchibergmann1995}, this type of large amplitude, rapid variability is now found in higher luminosity quasars
\citep{lamassa2015,macleod2016,ruan2016,runnoe2016,gezari2017,wang2018,yang2018,stern2018,macleod2018,trakhtenbrot2019}. Evidence from short timescale variability of both the broad emission lines and continuum \citep[e.g.,][]{lamassa2015}, a lack of change in optical reddening \citep{macleod2016} or X-ray column \citep{husemann2016}, echoed
mid-infrared variability (from warm dust) following optical
variability \citep{stern2018}, and spectro-polarimetry
\citep{hutsemekers2017} all imply that these objects undergo rapid changes in intrinsic accretion power on timescales much shorter than the $10^{4-6}$ year inflow time expected for a standard thin disk. In certain cases, the variability could be driven by a discrete event such as a tidal disruption flare
\citep{merloni2015}, microlensing \citep{bruce2017}, or a binary black hole interaction \citep{kim2018}. Generally, however, changing look events appear to be the rare excursions to large amplitude of the continuous and stochastic variability seen in ``hypervariable" \citep{rumbaugh2018} and ordinary \citep{kelly2009} quasars. 

Large amplitude, rapid variations in luminous quasars are difficult to explain in the context of standard accretion disk theory. Either the entire AGN optical SED must be powered by reprocessing of central UV or X-ray emission \citep{shappee2014,lawrence2018}, AGN are subject to disk instabilities possibly including fast state transitions \citep[e.g.,][]{noda2018,ross2018}, and/or inflow through the disk can be much more rapid than commonly assumed \citep{dexter2019}. 

These scenarios make markedly different predictions for spectral evolution during changing look events. Reprocessing in a thin disk should show coordinated variability at all wavelengths with blue leading red. The fractional variability amplitude should also decrease with increasing wavelength due to geometric dilution of the central flux and the larger variation in light travel time to the emission region at long wavelength \citep[e.g.,][]{kazanas2001,sergeev2005,cackett2007,shappee2014}. Rapid inflow should result in red wavelengths leading with long delay timescales corresponding to viscous evolution \citep{pringle1981}. Disk instabilities could produce either band leading, possibly accompanied by a change in SED shape due to changes in disk structure.

The quasar \sdssj\,at $z \simeq 0.456$ was recently identified as a ``turn-on" changing look object, in which the broad \ha\, and \hb\, lines increased in strength during a large change in continuum luminosity in 2017 \citep{wang2018}. This object is included in the Sloan Digital Sky Survey Reverberation Mapping (SDSS-RM, RMID 017) campaign that intensively monitors $849$ quasars with optical imaging and spectroscopy since 2014 \citep{shen2015}. We present extensive optical to X-ray photometric and spectral monitoring data of the source from 2009-2018 taken as part of the SDSS-RM campaign, including intensive SDSS spectral monitoring over the last five years (\autoref{sec:observations}), including from SDSS-IV \citep{blanton2017}. We use the optical continuum (\autoref{sec:continuum}) and broad emission line (\autoref{sec:lines}) evolution from this extensive data set to confirm that optical changing look quasars are powered by order of magnitude intrinsic variations in accretion power. From the line shape and luminosity evolution, we probe the low-ionization broad emission line region structure and physical conditions in a luminous quasar. We describe the difficulties in explaining these observations in terms of simple reprocessing or intrinsic disk emission models (\autoref{sec:discussion}).

\section{Observations}\label{sec:observations}

\sdssj\, was included in the SDSS-RM campaign \citep{shen2015}. SDSS-RM has observed 849 quasars in a 7 deg$^2$ field coinciding with the Pan-STARRS1 Medium Deep Field MD07 \citep[RA J2000=213.704, Dec J2000=+53.083,][]{tonry2012}. The monitoring included intensive spectroscopy with the Apache Point 2.5 meter SDSS telescope \citep{gunn2006}, along with accompanying photometry obtained from the CFHT and Bok telescopes. Pan-STARRS 1 photometry is also available from 2010-2013, along with GALEX observations and 2 XMM-Newton observations in 2017. We summarize the data set and reduction procedure used here.

\begin{table}[t]
\centering{}
\begin{tabular}{lcc}
\hline \hline 
\textbf{Band} & $\lambda_{\rm min}$ (\AA) & $\lambda_{\rm max}$ (\AA) \\
\hline
2700 & 2655 & 2720 \\
3050 & 3020 & 3100 \\
3500 & 3500 & 3600 \\
4000 & 4000 & 4050 \\
4500 & 4450 & 4550 \\
5100 & 5100 & 5200 \\
5500 & 5500 & 5600 \\
6000 & 6000 & 6100 \\
7000 & 6800 & 7000 \\
\hline
\end{tabular}
\caption{Rest frame continuum luminosity bands used here.}
\label{tab:1}
\end{table}

\begin{table*}[t]
\centering{}
\begin{tabular}{lccccccccc}
\hline \hline 
Line name & Blue cont. (\AA) & Line region (\AA) & Red cont. (\AA) & log$L_{\rm line,44}$ & $\eta$ & $\sigma_{\rm line,44}$ (km s$^{-1}$) & $p$\\ 
\hline
H$\alpha$ & 6326-6399 & 6399-6755 & 6755-6833 & 42.52 & $0.44 \pm 0.04$ & 3150 & $-0.21 \pm 0.02$\\ 
H$\beta$ & 4741-4762 & 4762-5027 & 5027-5050 & 42.03 & $0.35 \pm 0.05$ & 4370 & $-0.26 \pm 0.03$\\
H$\gamma$ & 4222-4271 & 4271-4401 & 4401-4452 & 41.33 & $0.81 \pm 0.06$ & 1920 & $-0.12 \pm 0.04$\\
MgII & 2717-2748 & 2748-2851 & 2851-2884 & 42.36 & $0.33 \pm 0.04$ & 1850 & $-0.12 \pm 0.02$\\
HeI & 5783-5850 & 5850-5931 & 5931-6000 & 40.73 & $0.93 \pm 0.09$ & -- & -- \\
HeII & 4475-4527 & 4527-4851 & 4851-4907 & 40.36 & $1.10 \pm 0.14$ & -- & --\\
\hline
\end{tabular}
\caption{Rest frame continuum and line fitting regions, line luminosities and responsivities ($\eta$), and line widths and breathing indices ($p$). The line luminosity and $\sigma_{\rm line}$ normalizations are scaled to a $2700$\AA\, luminosity $\nu L_\nu = 10^{44} \,\ergs$. We do not report line width information for the faint He lines where the measurements are unreliable.}
\label{tab:2}
\end{table*}

\subsection{SDSS Spectra}

The SDSS-RM spectra analyzed here were obtained with the BOSS spectrographs \citep{smee2013} between Jan 2014 and June 2018. We use data from 77 epochs, with a median cadence of only 4 days in 2014 and 16 days in the other years over 7 months of observing per year.

The exposure time was typically 2 hr, and the data were first processed by the BOSS pipeline, followed by a custom scheme to improve spectrophotometry and sky subtraction \citep[for technical  details on the SDSS-RM spectroscopy, see][]{shen2015}. The typical absolute spectrophotometric accuracy achieved is $\sim 5\%$. More detail on the spectroscopy data and analysis can be found in \citet{shen2018} and \citet{grier2019}. From the spectra, we estimate continuum luminosities as mean values over narrow bands (\autoref{tab:1}), avoiding clear emission lines in the time-averaged spectrum. We have not tried to model weak narrow emission lines or the pseudo-continuum from FeII. 

We measured emission line properties after subtracting a polynomial fit to the nearby continuum. The line flux is estimated by numerically integrating over the measured flux in each channel. The line width is charactered by both the FWHM and the square root of the second moment, $\sigma$. Both measurements are made directly on the data rather than using a model. The integration is performed over the spectral windows defined in \autoref{tab:2}. For \hb\, at earlier epochs, the line is very broad and blends with the [O~III]$\lambda 4959$ and $\lambda 5007$ doublet. We fit for and remove those forbidden lines separately before measuring the \hb\, line properties. The \oiii\, flux is constant to within an rms of 10\%, which is then an estimate of a typical flux error. There is no secular change in \oiii\, flux over time. Typical errors on the measured line width $\sigma_{\rm line}$, as measured by the distribution of their residuals when subtracting a power law dependence on luminosity (see \autoref{sec:lines}), are $\simeq 250$ km s$^{-1}$ for the reliably measured \ha, \hb, \mgii, and H$\gamma$ lines. We note that the PrepSpec code \citep[described in][]{shen2016} used for SDSS-RM reverberation measurements \citep{shen2016,grier2017,grier2019,homayouni2018} obtains \oiii\, flux measurements with $\lesssim 5\%$ rms.

We next subtract the host galaxy spectrum at each epoch using the template found from spectral decomposition by \citet{shen2015msigma}. The host galaxy contribution exceeds that of the AGN for rest frame $\lambda > 4000$\AA\, in 2014-2015. Finally, we convert the observed frame flux to the emitted monochromatic luminosity $\nu L_\nu$ assuming luminosity distances based on a WMAP 9 year cosmology \citep{hinshaw2013} as implemented in \texttt{astropy} \citep{astropy} ($H_0 = 69.3\, \rm km\, \rm Mpc^{-1}\, \rm s^{-1}$, $\Omega_{\rm M} = 0.287$, $\Omega_\Lambda = 0.713$).

\subsection{SDSS-RM photometry}

Photometric observations of \sdssj\ were obtained in both the $g$ and $i$ bands \citep{fukugita1996} using the Steward Observatory Bok 2.3m telescope on Kitt Peak and the 3.6m Canada-France-Hawaii Telescope (CFHT) on Mauna Kea. They cover roughly the same monitoring period as the SDSS-RM spectroscopy in 2014 with a cadence of about 2 days in 2014 and more sparse in the following years. The Bok observations were obtained with the 90Prime instrument \citep{williams2004}, which has a 4k $\times$ 4k CCD with a plate scale of 0.\arcsec45 pixel$^{-1}$ and a field of view of 1\degree~$\times$~1\degree. The CFHT observations were taken with the MegaCam instrument \citep{aune2003}, which has a pixel scale of 0.187\arcsec and also has a 1\degree~$\times$~1\degree\ field of view. Details of these observations and their processing will be provided by K. Kinemuchi et al. (in preparation). 

The data were processed via pipelines for their respective instruments. Light curves were produced using the image subtraction method of \cite{alard1998, alard2000} using the publicly available software ISIS. This method aligns the images and then subtracts each individual epoch from a reference frame that is produced from a subset of the highest-quality images. Before subtraction, ISIS alters the PSF of the reference frame to match that of the individual epoch. This process results in a subtracted image that contains the residual flux, on which a PSF-weighted aperture is placed to measure the flux and produce a residual-flux light curve.

For 2014-2017, we also produce a version of the merged photometry and errors between SDSS, Bok, and CFHT. We use the CREAM code and procedure described in \citet{starkey2016}. CREAM models the light curve allowing an additive offset, multiplicative scaling factor, and transfer function for each band/instrument/telescope. These parameters are optimized using a Markov Chain Monte Carlo fitting process. The rescaled $g$ and $i$ band light curves are placed on the same scale and merged into a single continuum light curve. The uncertainties are further rescaled based on the variance between neighboring epochs. The impact of this rescaling is to increase the error bars, typically by factors $\simeq 2-3$ \citep{grier2019}. In this procedure we found a systematic offset of the CFHT data in 2016, likely due to a switch to different filters, and removed them from the merged light curve.

\subsection{PS1 photometry}

We used PS1 light curves of \sdssj\, in the $g$,$r$,$i$,$z$,$y$ bands from the Medium Deep Field Survey MD07 field \citep{tonry2012} covering the time period 2010-2013. The light curves are taken from \citet{shen2019} and details of the photometric calibration are described in \citet{schlafly2012} and \citet{magnier2013}. We have further removed flagged data (where there is no measured error) and a few obvious outliers (measurements $> 4$ standard deviations away from the median of that year).

\subsection{GALEX photometry}

A GALEX NUV light curve was obtained using data from the GALEX Time Domain Survey \citep{gezari2013}.  We performed aperture photometry on each GALEX epoch at the position of the nearest source to the SDSS position within 3 arcsec in a deep coadd of all the GALEX images.  We use an aperture radius of 6 arcsec, and apply an aperture correction for the energy enclosed, and a zeropoint for the AB magnitude system.  The photometric error was determined for each epoch empirically from the standard deviation of a control sample of sources between epochs as a function of magnitude, following \citet{gezari2013}. We detect the quasar in 19 GALEX epochs between May 2009 and March 2011.

\subsection{XMM-Newton}

\sdssj\, was observed by XMM-Newton twice, in June 2017 (OBSID=0804271301) and November 2017 (OBSID=0804271801). The EPIC net counts in the 0.5-10 keV band were 3789 and 420 for the two observations with exposure times of 17ks and 5.4ks. We extracted the
PN, MOS1, and MOS2 spectra and fitted them with an absorbed power-law plus a Compton reflection component (``pexrav"). We assumed a Galactic absorbing column density $N_H=1.1\times10^{20} \hspace{3pt} \rm cm^{-2}$. Between the two
observations (5 months separation), the $2-10$ keV spectral shape remained an
unobscured power-law. The photon index, defined by $N(E) \propto E^{-\Gamma}$, of $\Gamma = 2.15 \pm 0.05$ remained constant between observations, while the flux dropped by a factor of 3.6. The unabsorbed rest-frame 2-10 keV luminosities of the two observations are $L_X = (3.09 \pm 0.17) \times 10^{44}\,\ergs$ and $(8.6 \pm 1.0) \times 10^{43}\,\ergs$. Fitting the two spectra simultaneously gives a $90\%$ upper limit for any absorbing column of $N_H =9 \times 10^{19} \,\rm cm^{-2}$. 

\begin{figure*}
    \centering
    \begin{tabular}{ll}
    \includegraphics[width=\columnwidth]{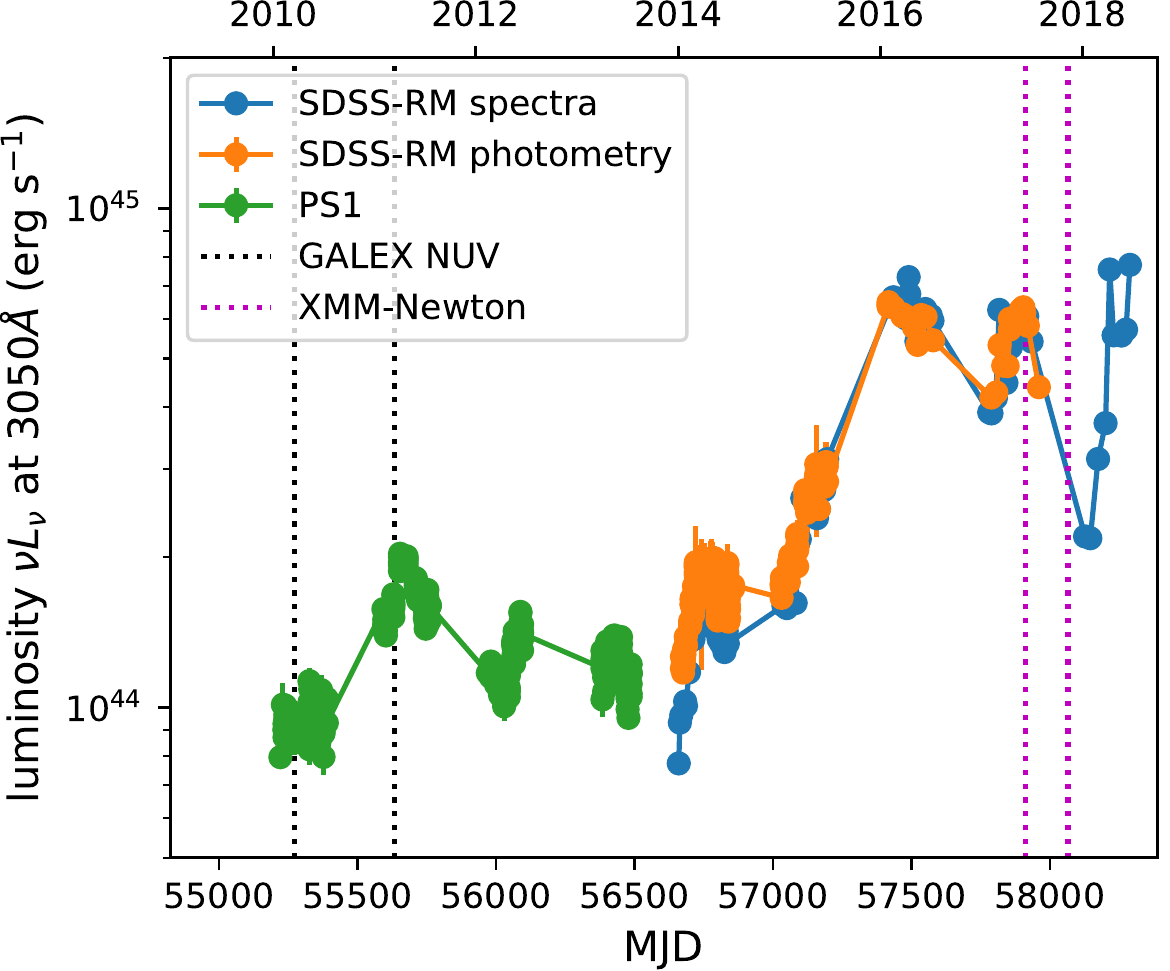} &
    \includegraphics[width=\columnwidth]{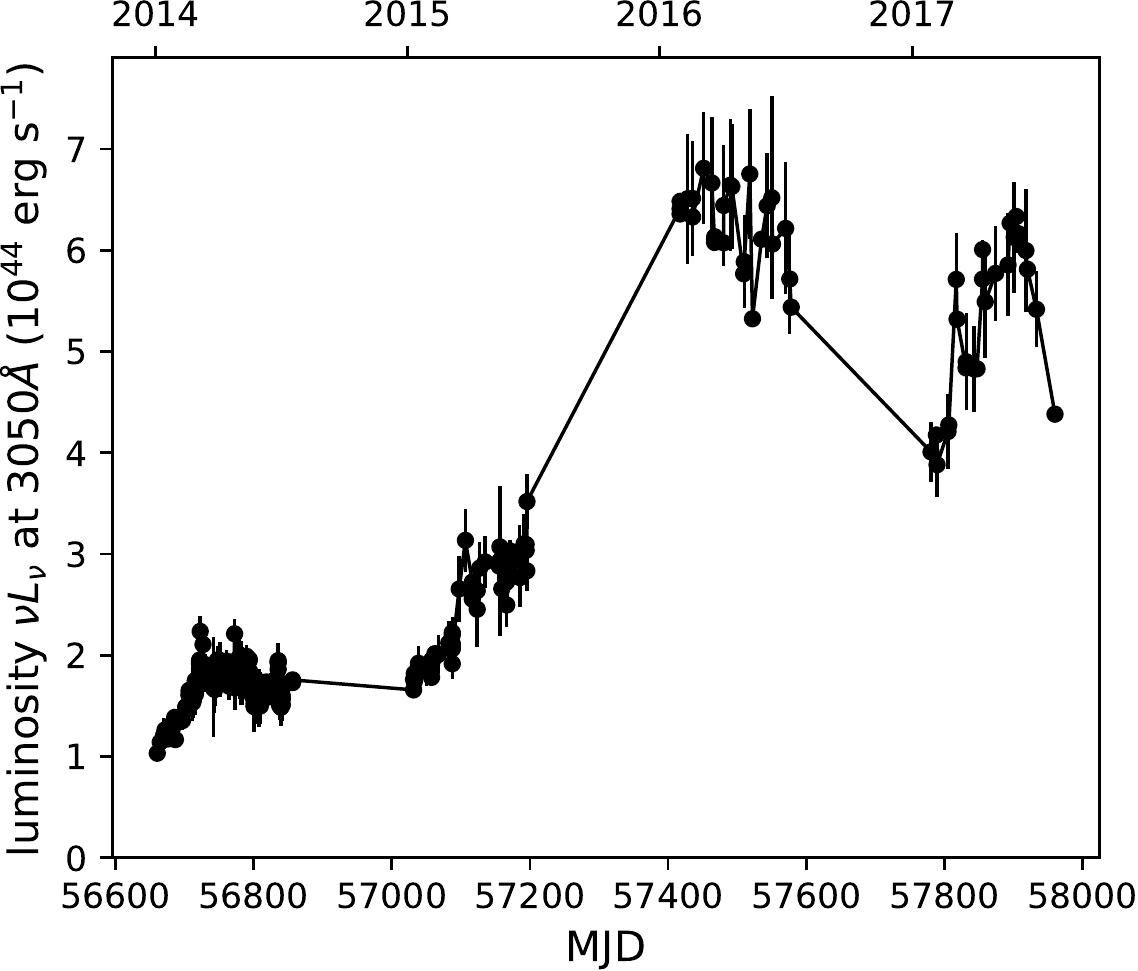}\\
    \end{tabular}
    \caption{Continuum optical light curve for \sdssj\, from
      2010-2018 (left). The green and orange points are luminosities derived
      from Pan-STARRS PS1 and SDSS-RM $g$-band photometry, while the
      blue points are continuum measurements in the rest frame $3050$\AA\, band
      defined in \autoref{tab:1}. The dotted curves mark the extrema used to measure SEDs including GALEX NUV data, and the epochs of the XMM-Newton observations. The light curve shows continuous,
      rapid, large-amplitude variability. Features include a prominent rise and fall by a factor of 2 from 2010-2012, followed by a dramatic rise by almost an order of magnitude over a year from 2015-2016, and a sharp dip and final peak in 2018. In 2018 the optical luminosity increased by a factor of 3-6 in only 6 months. A zoom in of the 2014-2017 merged light curve is shown in the right panel on a linear scale.}
    \label{fig:gband_combine}
\end{figure*}

\begin{figure}
    \centering
    \includegraphics[width=\columnwidth]{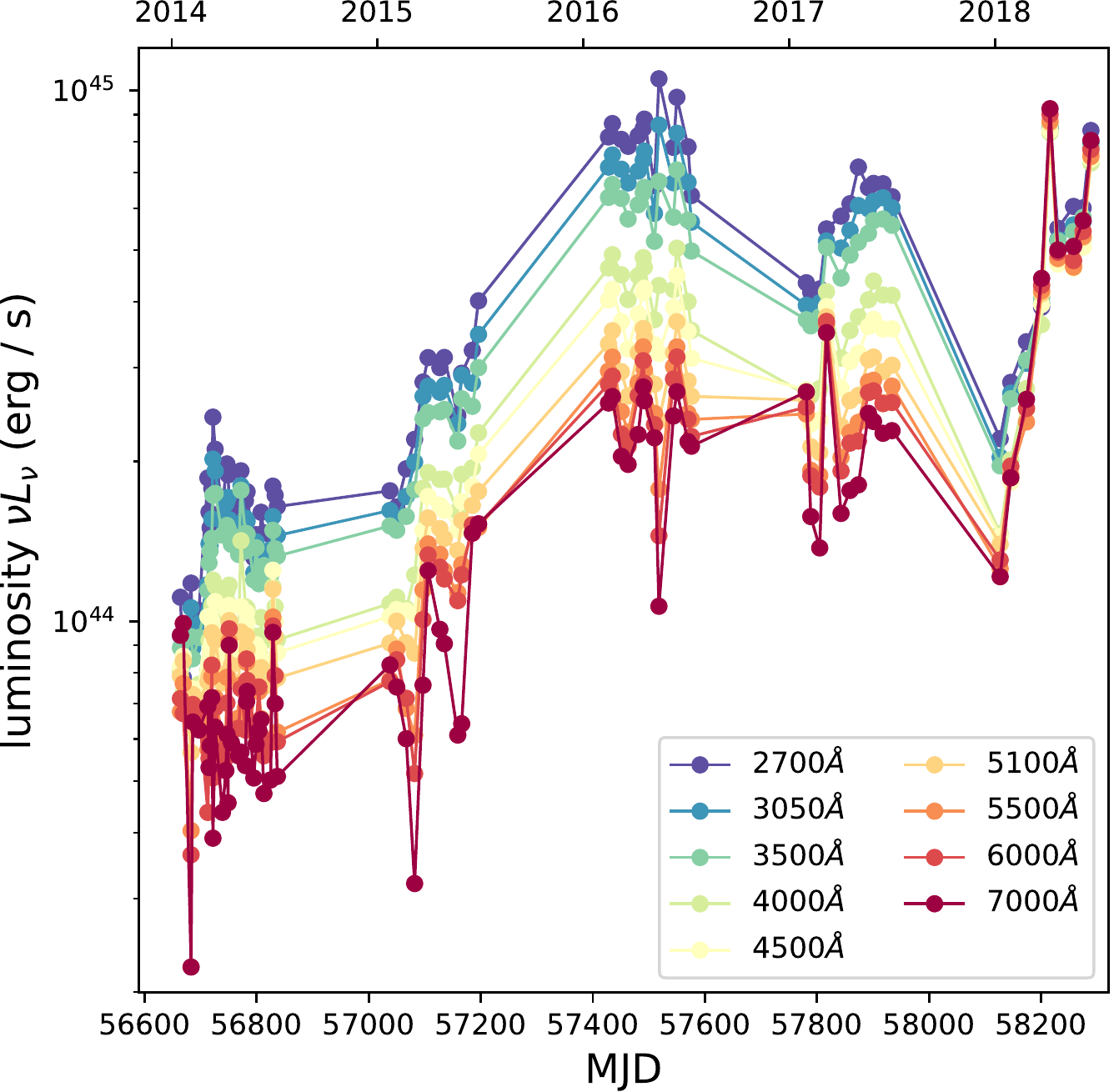}
    \caption{Host galaxy subtracted, rest frame multi-band continuum light curves measured from the SDSS spectra. During the initial large rise the continuum maintains a blue shape. In the later drop and sharp rise in 2018 it suddenly reddens. At the final peak, the SED is completely flat across the observed optical range. The host galaxy dominates in the red bands ($\lambda \gtrsim 4000$\AA) in 2014-2015 when the AGN optical luminosity is low \citep[$L \lesssim 10^{44}\,\ergs$,][]{shen2015msigma}.}
    \label{fig:multiband}
\end{figure}

\section{Rapid, large amplitude continuum variations}
\label{sec:continuum}

\sdssj\,shows continuous, large amplitude optical variability over the past $\simeq 10$ years (left panel of \autoref{fig:gband_combine}). The initial Pan-STARRS monitoring showed a factor $\simeq 2-3$ increase (near MJD 55500) from the luminosity inferred from the 2004 SDSS spectrum. A drop and similar rise occurred again during the high cadence SDSS-RM monitoring in 2014 (near MJD 56600). Subsequently the luminosity increased by a factor $\simeq 10$ over roughly a year period. It has since stayed high, but with large excursions e.g. a factor $\gtrsim 2$ drop in 2017 in all bands (near MJD 58100) followed by a rise by a factor of $\simeq 3-6$ within the 6 month observing window during 2018. The period from 2014-2017 is shown in more detail, with estimated photometric uncertainties from the CREAM modeling and merging, in the right panel of \autoref{fig:gband_combine}.

The variations are highly correlated and nearly simultaneous across the measured optical band. This is shown for the SDSS-RM spectra in \autoref{fig:multiband}. A detailed continuum lag analysis for the high cadence data found that the $g$-band leads 
the $i$-band by $\simeq 3-6$ days \citep{homayouni2018}. The lower cadence of the later light curve precludes a continuum lag measurement, but continues to show rapid and highly correlated variations consistent with the high cadence monitoring.

Until 2018, the fractional amplitude changes are similar across all optical bands. The blue bands increase somewhat more during the first dramatic rise with $\nu L_\nu$ consistent with a power law of slope $1.2-1.4$, similar to the value of $4/3$ expected for a thin disk. The subsequent flattening, drop, and dramatic rise in 2018, by contrast, show dramatic reddening of the SED. By the end of observations in 2018 the SED was completely flat in $\nu L_\nu$. Compared to the previous peak at similar luminosity $\simeq 10^{45}\,\ergs$, the reddest bands increased in luminosity by a factor $\gtrsim 3$ while the bluest band slightly decreased. For the estimated RM black hole mass measured from \hb\ of $M \simeq 8 \times
10^{8} M_{\odot}$ \citep{grier2017}, the optical luminosity varies between $L_{\rm opt}/L_{\rm Edd} \sim 0.2-2\%$, similar to other hypervariable quasars which tend to have low Eddington ratios \citep{rumbaugh2018}. The GALEX near-UV light
curve shows similar evolution to that of the $g$-band from PS1 (\autoref{fig:gband_combine}) but with a slightly higher fractional rms amplitude.

\begin{figure}
    \centering
    \includegraphics[width=\columnwidth]{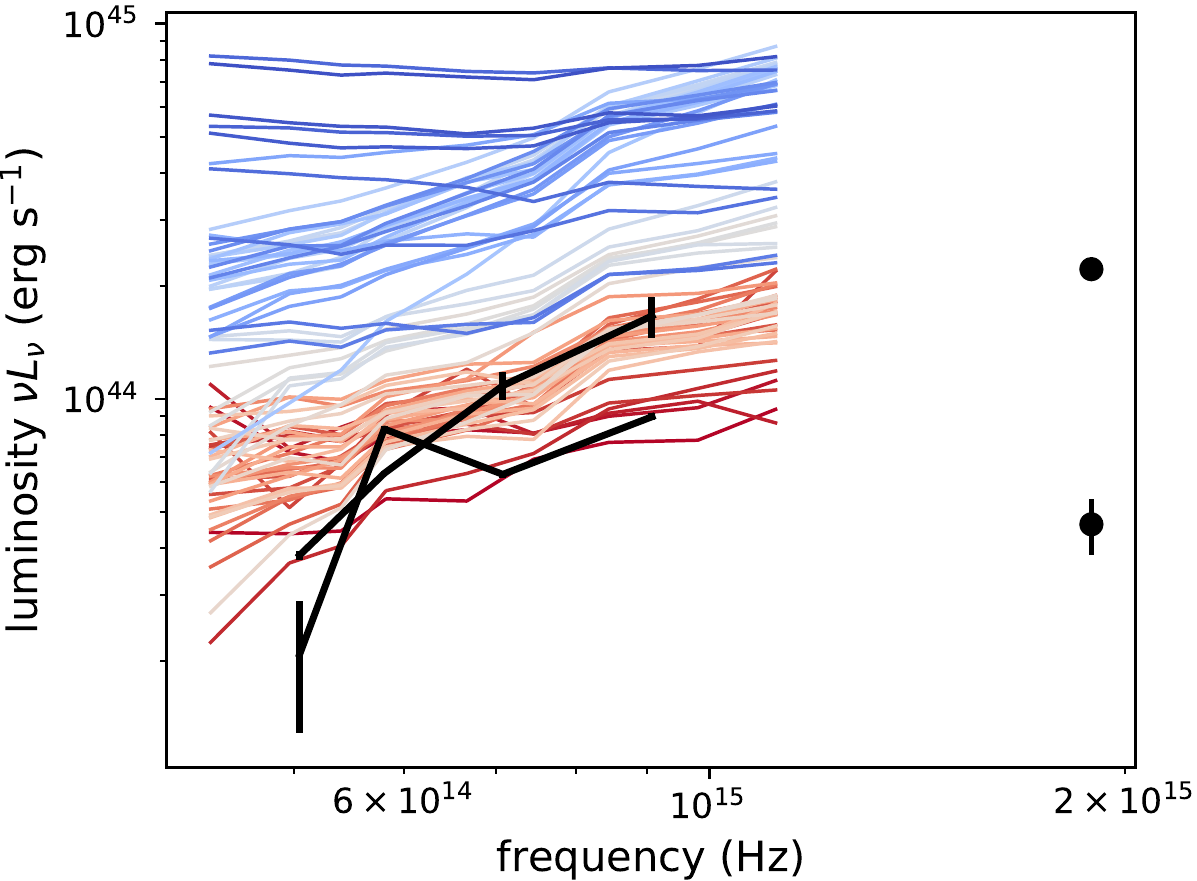}\\
    \includegraphics[width=\columnwidth]{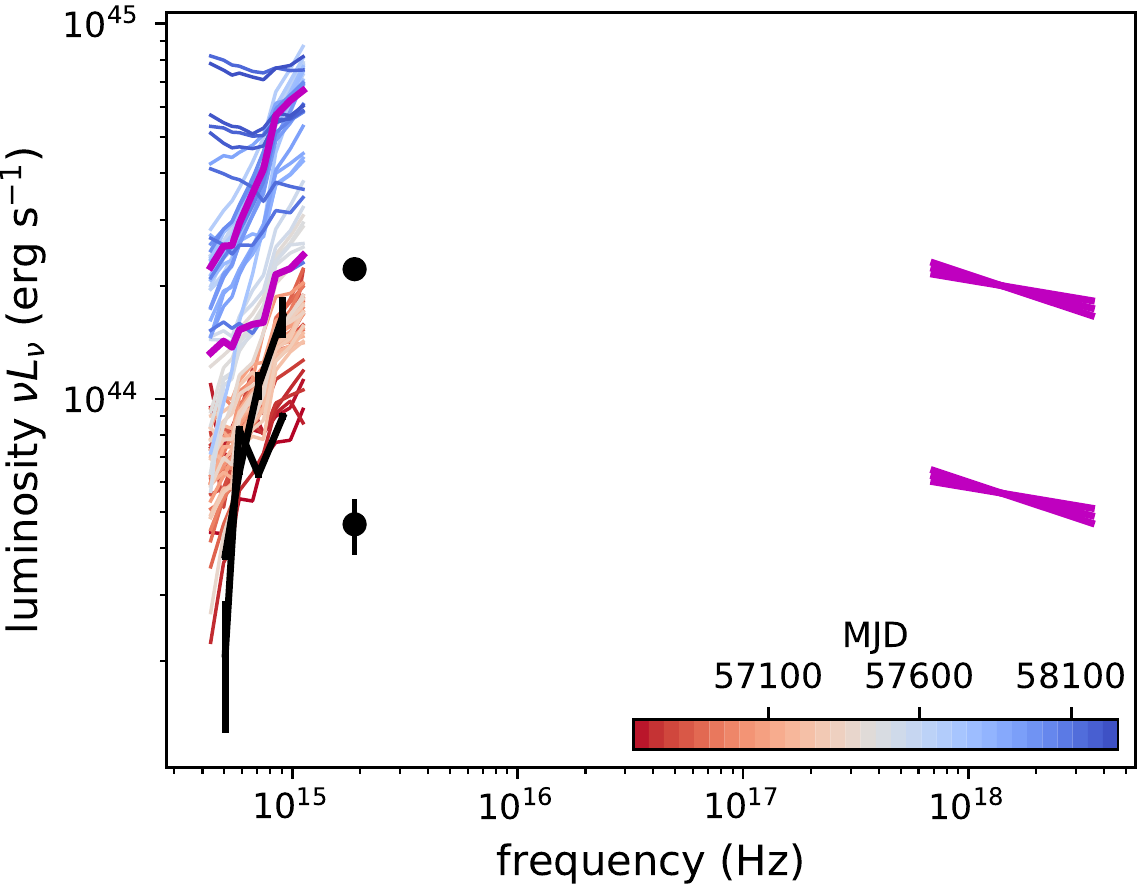}
    \caption{Host galaxy subtracted, rest frame, optical to X-ray SEDs. The black curves and points show quasi-simultaneous PS1 and GALEX photometry. The magenta curves show quasi-simultaneous continuum measurements from SDSS spectra along with XMM-Newton data. The colored curves are measurements from SDSS-RM spectra going from early (red) to mid (gray) to late (blue) epochs. The optical SEDs are consistent with power laws. In the lower state, the SED peak appears to be redwards of the GALEX NUV point. In the GALEX higher state and most of the SDSS-RM spectra, the peak is bluewards of our observing band. Where measured, the X-ray luminosity is relatively high but well below that of the bluest optical band. During the dramatic rise in 2018, the SED reddens to become completely flat across the optical band.}
    \label{fig:sed}
\end{figure}

Sample host-subtracted SEDs are shown in \autoref{fig:sed}. We choose two epochs for the PS1/GALEX data (black) corresponding to low (2009) and high (2011) luminosity states. Continuum luminosities measured from the later SDSS spectra are shown as colored lines. Until 2018 (MJD $\simeq 58100$), the SED rises
beyond our bluest optical band, except maybe in the low state of the PS1/GALEX monitoring. This is broadly consistent with a spectral break near $1000$\AA\, as seen in quasar composite spectra \citep[e.g.,][]{shull2012}. For the estimated mass of $\sim 10^9 M_{\rm sun}$ and observed $L_{\rm opt}$, thin accretion disk models predict a peak in $\nu L_\nu$ at $\simeq 10^{15}$ Hz, broadly consistent with a peak location in between the PS1 $g$ and GALEX NUV bands. In many epochs the continuum slope is comparable to the expected $\nu L_\nu \sim \nu^{4/3}$, while notably in 2018 it is much flatter. The X-ray luminosities measured by XMM are a factor of $\simeq 3$ lower than quasi-simultaneous measurements in our bluest SDSS band. The resulting spectral index $\alpha_{\rm OX}
\simeq -1.2$. The source is mildly X-ray bright, roughly $1\sigma$ off the relation for ordinary quasars
\citep{steffen2006,lusso2010} and similar to other hypervariable ones \citep{rumbaugh2018,collinson2018}. We do not find evidence for a change in
$\alpha_{\rm OX}$ between the two epochs where the X-ray luminosity varies by a factor $\simeq 4$.

\begin{figure*}
    \centering
    \includegraphics[width=0.9\textwidth]{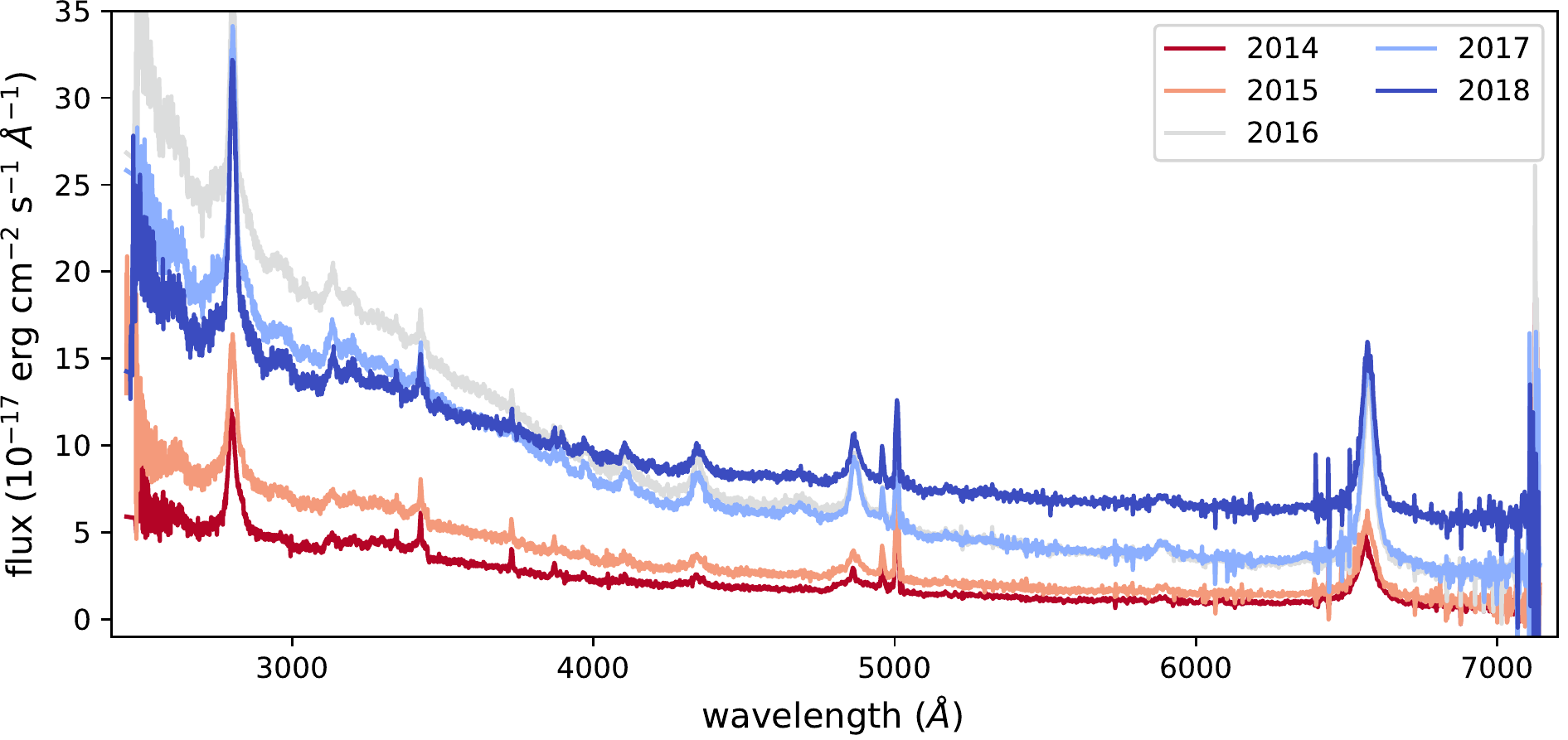}
    \caption{Host galaxy subtracted spectra averaged over each observing season. The continuum luminosity and shape change dramatically between years.}
    \label{fig:spectra_years}
\end{figure*}

\begin{figure*}
    \centering
    \begin{tabular}{ll}
    \includegraphics[width=0.48\textwidth]{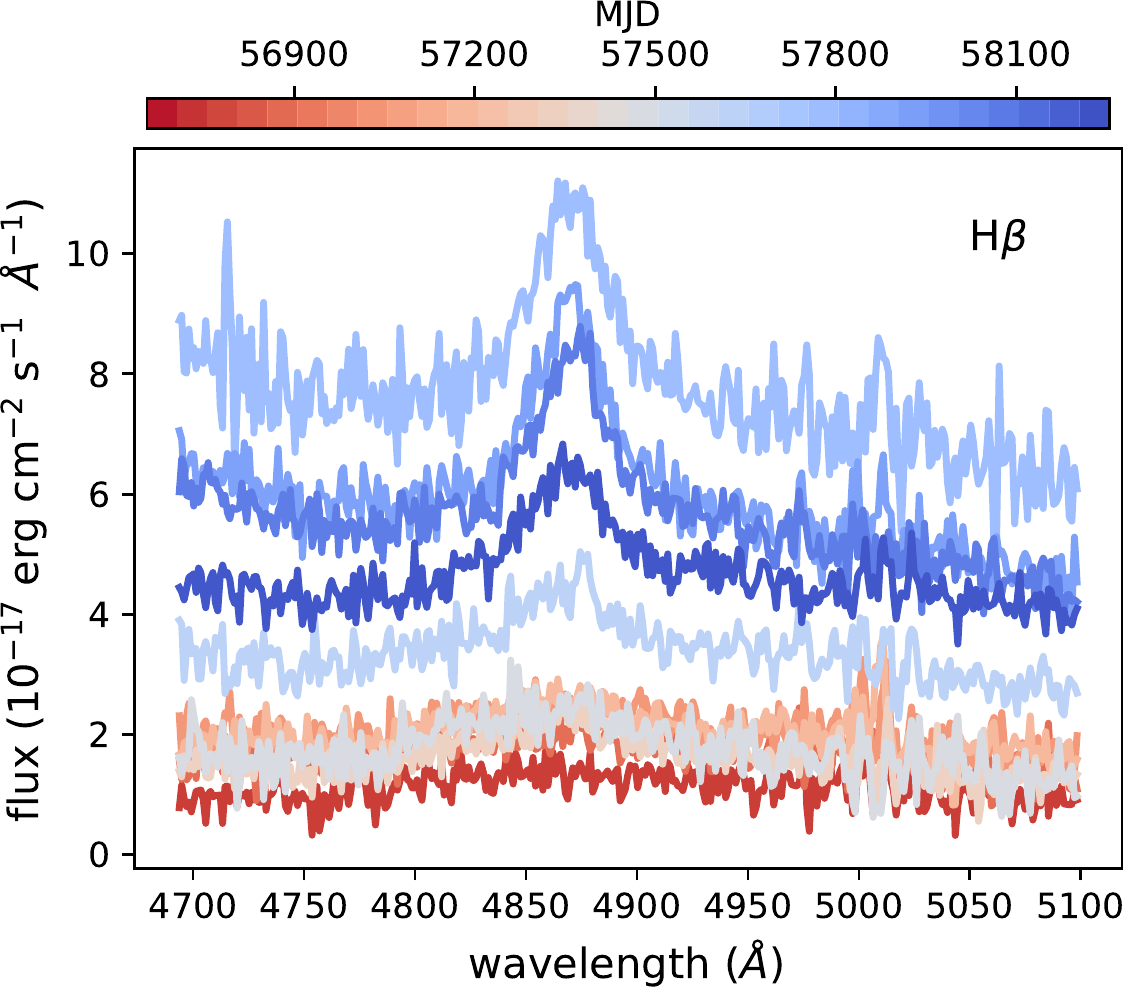} &
        \includegraphics[width=0.48\textwidth]{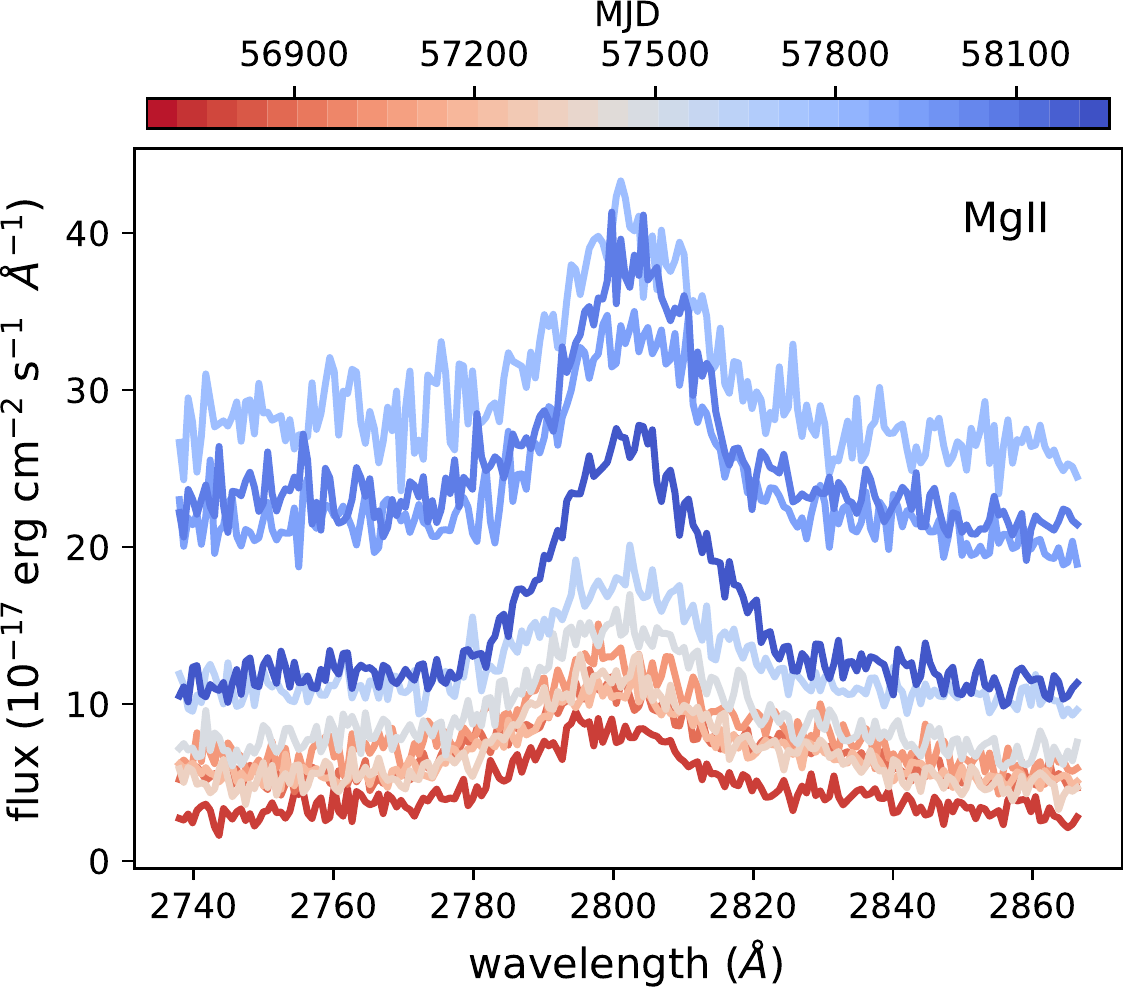}
    \end{tabular}
    \caption{Sample host galaxy and narrow line subtracted \hb\, (left) and \mgii\, (right) profiles from early (red) to blue (late) epochs. The line becomes significantly narrower at late times where the continuum luminosity is higher. The shape also changes, with the core brightening more than the wings.}
    \label{fig:lines_sample}
\end{figure*}

\begin{figure}
    \centering
    \includegraphics[width=0.48\textwidth]{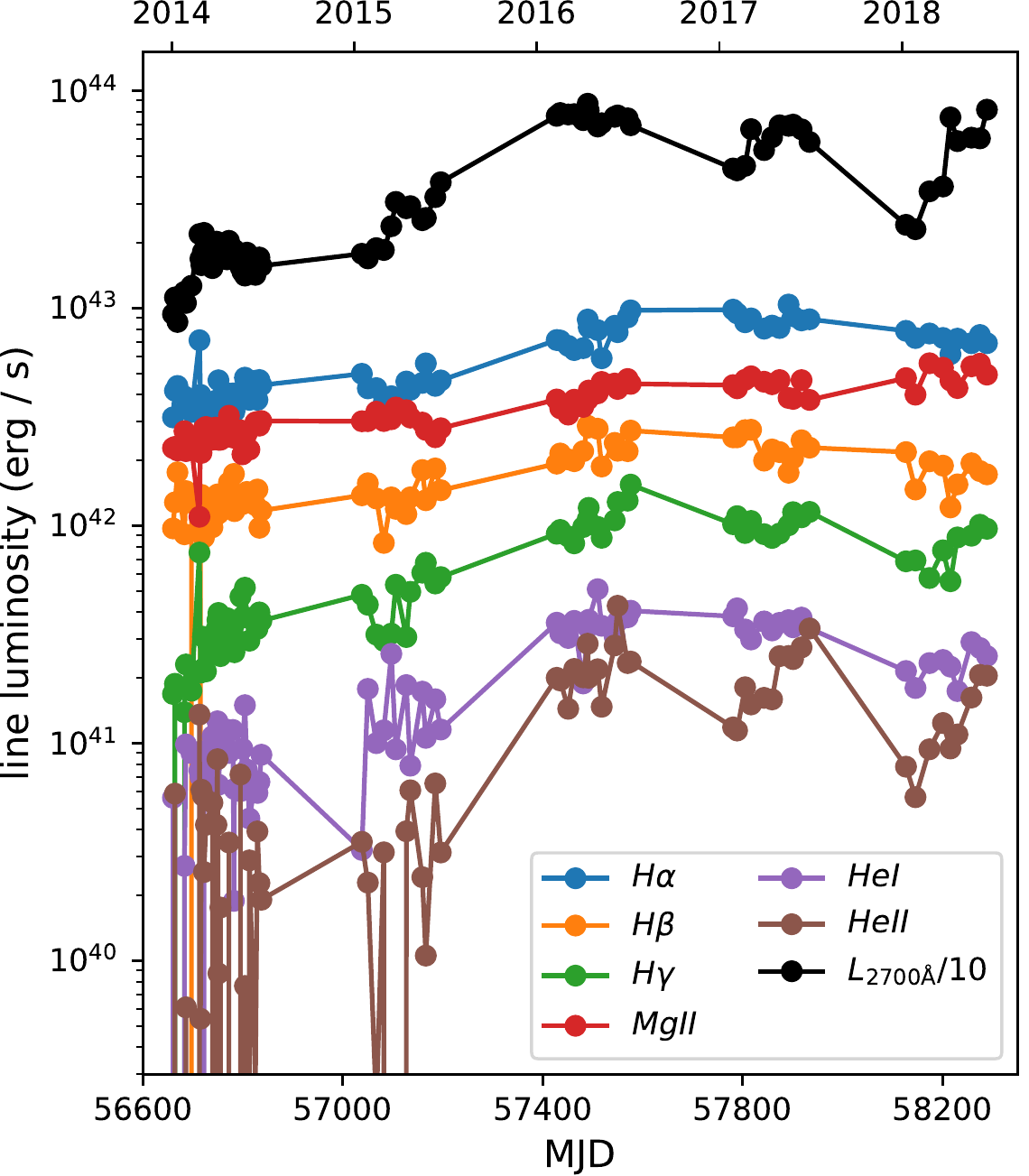}
    \caption{Continuum and broad emission line light curves measured from the SDSS spectra. All lines increase in luminosity during the large rise in the optical continuum. The He lines show a stronger response than the Balmer and \mgii\, lines.}
    \label{fig:lines_lc}
\end{figure}

\begin{figure*}
    \centering
    \begin{tabular}{cc}
    \includegraphics[width=0.45\textwidth]{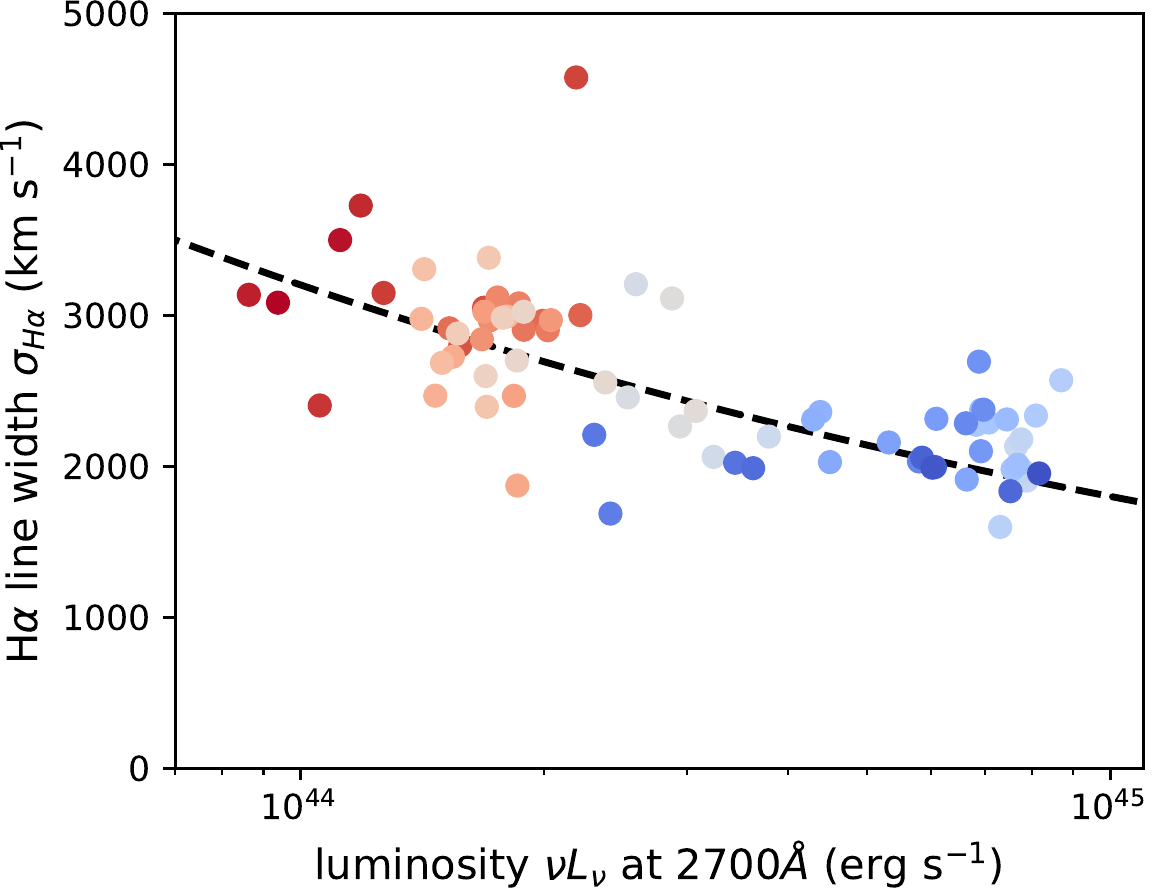} &
    \includegraphics[width=0.45\textwidth]{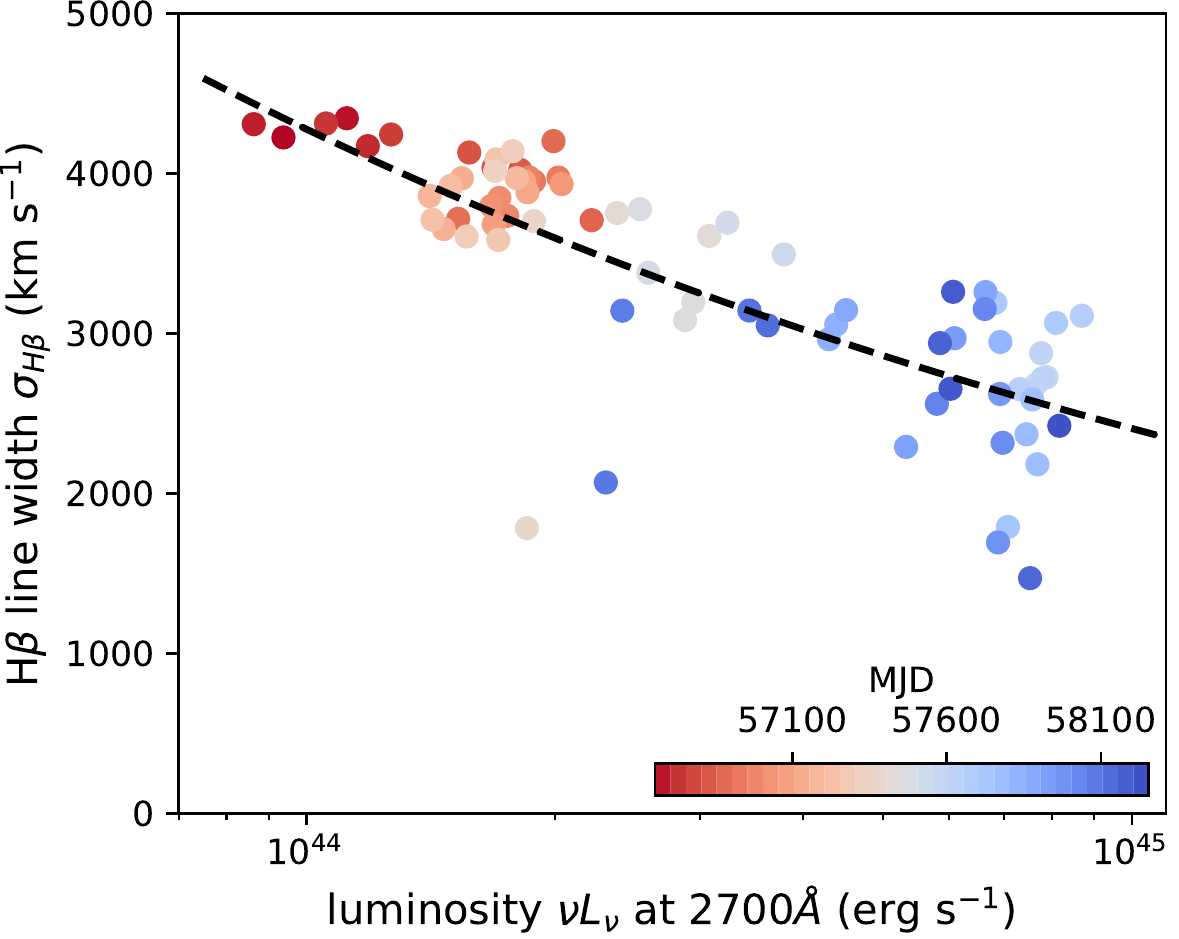}\\
    \includegraphics[width=0.45\textwidth]{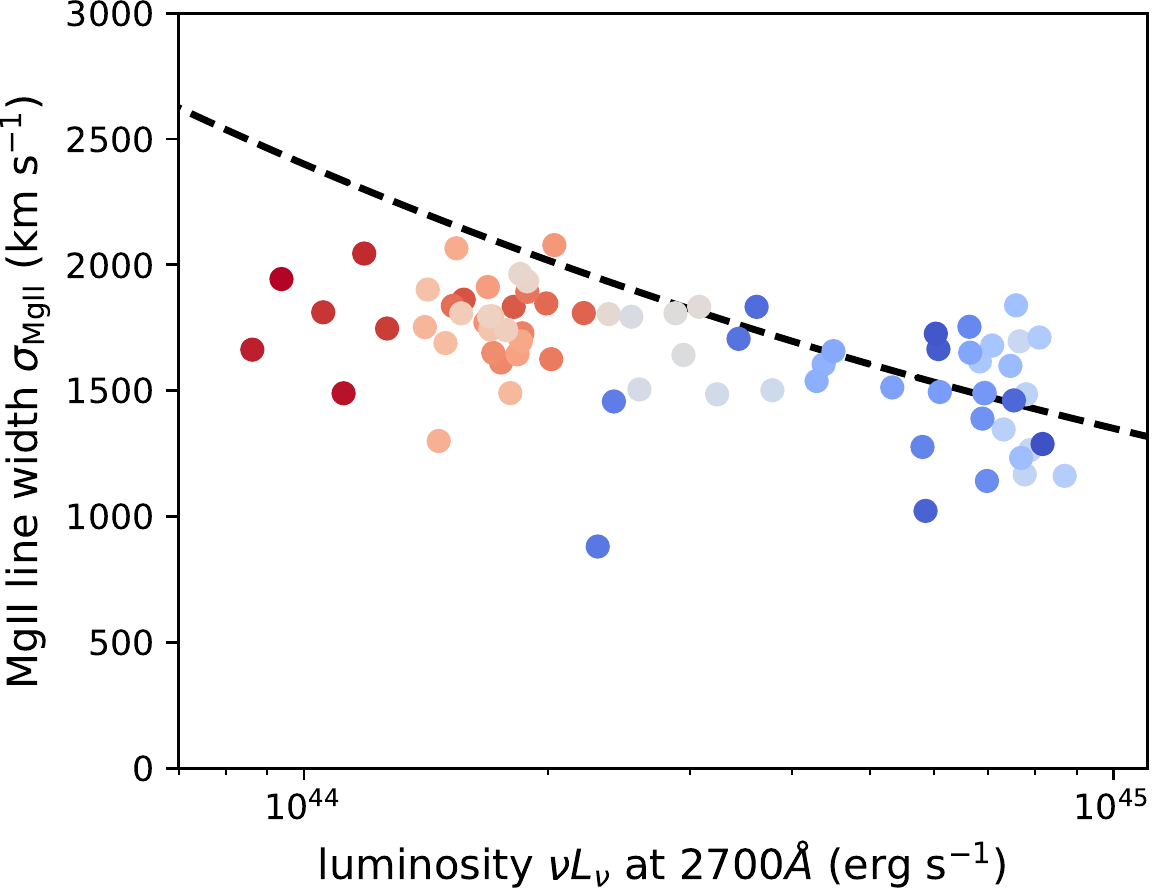} &
    \includegraphics[width=0.45\textwidth]{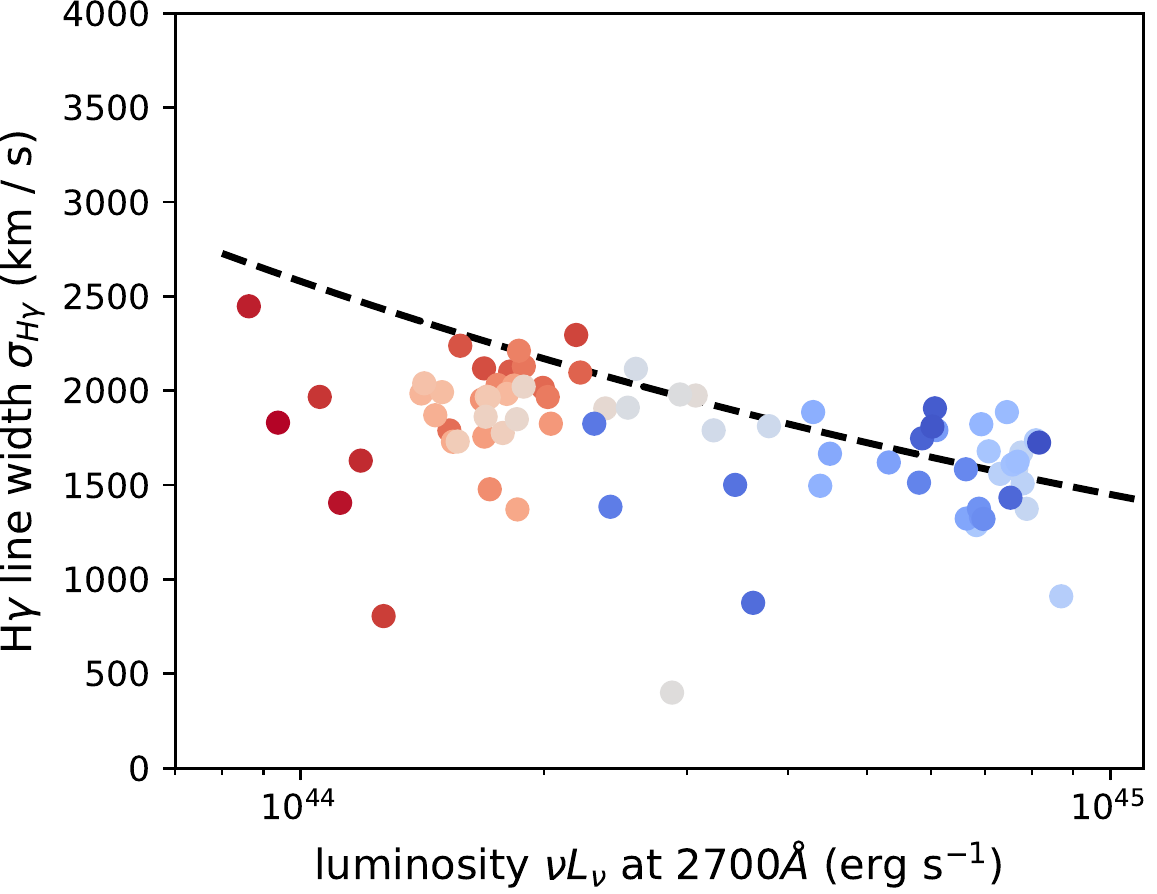} \\
    \end{tabular}
    \caption{Line width measurements for \ha, \hb, \hg, and \mgii\, as a function of continuum optical luminosity. Point color indicates observational epoch (red early to blue late). The \ha\, and \hb\, widths are clearly anti-correlated with luminosity. The same trend is found for \mgii\, and H$\gamma$, but with a weaker dependence. The dashed curve is a relation $\sigma \sim L^p$ with $p=-1/4$, expected if the BLR radius $R_{\rm BLR} \sim L^{1/2}$ and radial velocity $v \sim R^{-1/2}$. The consistency of this trend with the data demonstrates the expected ``breathing" of the BLR in response to continuum variations.}
    \label{fig:blr_breathing}
\end{figure*}

\begin{figure}
    \centering
    \includegraphics[width=\columnwidth]{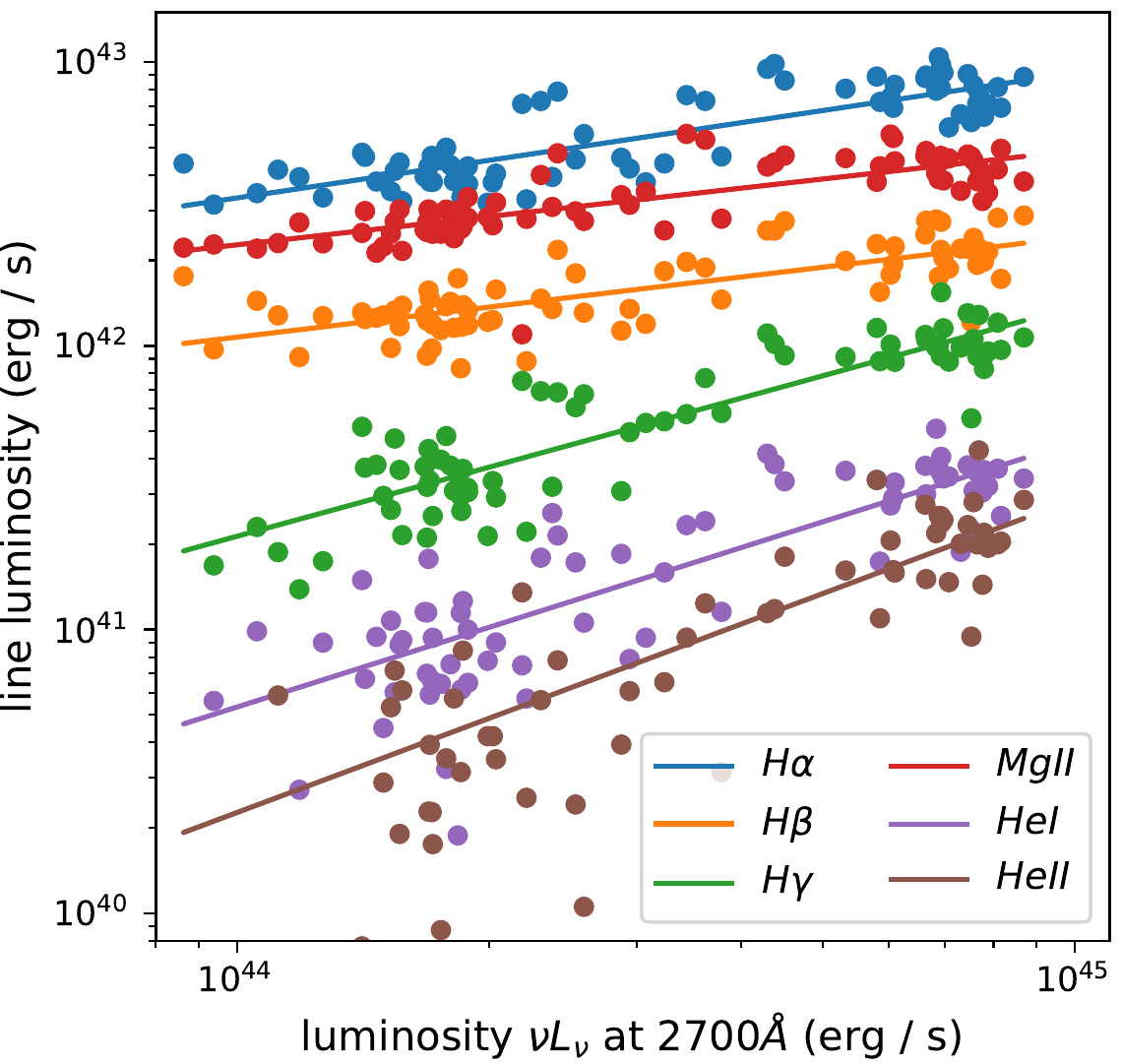}
    \caption{Line vs. $2700$\AA\, rest frame continuum luminosity for all broad emission lines measured. The solid lines are power law fits. Line luminosity is correlated with that of the continuum, but with a sub-linear slope especially for the low-ionization Balmer and \mgii\, lines.}
    \label{fig:blr_response}
\end{figure}

\section{Broad emission line response}
\label{sec:lines}

Host galaxy subtracted spectra averaged by year are shown in figure \autoref{fig:spectra_years}. All observed broad emission lines respond to the dramatic changes in the optical continuum. Example line profiles are shown for \hb\, and \mgii\, in  \autoref{fig:lines_sample}, with colors coded by time from red (early) to blue (late). In both cases the lines brighten dramatically, especially in their lower velocity cores. The classification for \hb\, changes from Type 1.8 to 1, which led to the classification of \sdssj\, as a turn-on changing look quasar \citep{wang2018}. Nonetheless, broad \mgii\, and \ha\, emission are detected at all times. We also note that \autoref{fig:spectra_years} provides evidence for increased O~III $\lambda 3133$ emission in high flux states, which could be produced by Bowen fluorescence \citep{trakhtenbrot2019bowen}. The similarly excited N~III $\lambda 4640$ line is weak, which might suggest the metallicity in this object is not supersolar.

\autoref{fig:lines_lc} shows line luminosity vs. time for all broad emission lines detected: the Balmer lines \ha, \hb, H$\gamma$ as well as \mgii, \hei\, and \heii. The general rising trend is clear in all cases. The variability amplitude of the Balmer lines and \mgii\, is much smaller than that of the He lines which more closely follow the changes in the optical continuum. The host galaxy subtraction leads to large systematic uncertainty in the \hei\, and \heii\, measurements in 2014-2015 ($L_{\rm line} \lesssim 10^{41} \,\ergs$). Many of those points should be interpreted as upper limits. 

For the more luminous Balmer lines and \mgii\, we can also measure changes in the line widths (but not for \hei\, and \heii). Line widths for \ha, \hb, \mgii, and \hg\, as a function of optical luminosity are shown in \autoref{fig:blr_breathing}, with no correction made for the $\simeq 40$ day light travel time delay to the BLR \citep{grier2017}. We show the results based on $\sigma$, but the results based on FWHM are similar. The \ha\, and \hb\, lines become significantly narrower at high luminosity. Similar behavior is seen in \mgii\, where the line core visibly narrows at higher luminosity (\autoref{fig:lines_sample}), but with a weaker luminosity dependence. This is consistent with the behavior seen in larger samples of both SDSS-RM and hypervariable quasars, where \mgii\, shows little or no response to continuum changes \citep{sun2015,wang2019}. The H$\gamma$ trend is also weaker, but the inferred line widths are uncertain at low luminosity due to the low signal-to-noise of the line.

We interpret the observed decrease in line width as the expected ``breathing" of the BLR, where the emission peak moves to larger radius from the black hole at higher luminosity \citep{baldwin1977,koratkar1991,cackett2006}. The dashed black curves in \autoref{fig:blr_breathing} show the trend of $\sigma \propto L^p$ with $p=-1/4$, expected if the BLR radius $R_{\rm BLR} \propto L^{1/2}$ and $\sigma \propto R^{-1/2}$ \citep[and assuming a constant SED shape, e.g.,][]{peterson2002}. Those simple scalings are consistent with the data for \ha\, and \hb. The \mgii\, line width is also well constrained and shows a breathing effect, but with a shallower index (best fitting values and uncertainties are listed in \autoref{tab:2}). In all cases the relevant velocity traced by the line width is falling with distance from the black hole. In other hypervariable quasars, the \mgii\, line width is independent of continuum luminosity \citep{yang2019}. The FWHM shows similar breathing behavior, consistent with previous work \citep[e.g.,][]{shen2013,zhang2018}.

The line and continuum optical luminosities are correlated in all cases (\autoref{fig:blr_response}). The large change in continuum luminosity provides a lever arm for measuring the responsivity $\eta$ for each emission line, defined here as $L_{\rm line} \propto L_{2700\text{\AA}}^\eta$. Best fits of that form are indicated as solid lines in the figure, and the best-fitting values and uncertainties are given in \autoref{tab:2}. The Balmer lines and MgII show flatter, sub-linear correlations (smaller $\eta$), while the He line response is nearly linear. These findings are in excellent agreement with the predictions of photoionization models \citep[e.g.,][]{korista2004}. Among the Balmer lines, the strongest response is in $H\gamma$, consistent with those models. The $H\beta$ response is weaker than expected.

\section{Discussion}
\label{sec:discussion}

We have analyzed optical to X-ray photometric and spectroscopic data of \sdssj, a hypervariable, changing look quasar in the SDSS-RM sample \citep[identified as such by][]{wang2018}. The large number of SDSS spectra as well as PS1 and GALEX NUV photometry and XMM-Newton spectroscopy allow a detailed study of  dramatic changes in a luminous quasar.

Past SDSS-RM measurements of \sdssj\, found observed frame time delays between the optical $g$-band and the \hb\, line and $i$-band continuum emission regions of $\simeq 30-50$ and $\simeq 3-6$ days, respectively. The measured BLR lag and line width leads to a black hole mass estimate of $M \simeq 8\times10^8 M_\sun$ \citep{grier2017}. The continuum reverberation timescale implies a propagation speed of $\gtrsim 0.5$c for a thin accretion disk \citep{homayouni2018}, consistent with light travel time delay where the optical variability is driven by reprocessed UV or X-ray emission.

We have shown that the source increased in brightness by a factor of 10 within about a year from 2015-2016. The variations are coordinated across the optical spectrum, with similar amplitudes at all bands. The SED is peaked towards the blue during the first rise, but remarkably turns red at the latest maximum in 2018, with roughly equal luminosity in all observed optical bands. 

First, we use the broad emission line response to the changing optical continuum to demonstrate that the changes are due to intrinsic variations in accretion power. Then we assess physical scenarios proposed to account for quasar hypervariability. We consider ``extreme reprocessing," where the entire optical spectrum is driven by reprocessing of a varying central source, and accretion instabilities where the observed changes are caused by rapid evolution of the disk structure.

\subsection{Intrinsic variations in accretion power}

There are many lines of evidence suggesting that optical changing look AGN are powered by intrinsic changes in central engine luminosity, rather than by changes in obscuration \citep[as is commonly seen in X-rays, e.g.,][]{risaliti2002,matt2003,yang2016}. The changes occur over a few years. To cover both the BLR and optical  emission region, an obscurer would need both a super-orbital velocity and large covering factor \citep[e.g.,][]{lamassa2015,runnoe2016}. When observed, the mid-infrared luminosity also varies in response to to the changing continuum \citep{sheng2017,stern2018}. X-ray data show no sign of an increase in column density associated with dramatic dimming of the continuum \citep[e.g.,][]{husemann2016}. Spectro-polarimetry of ``turn off" objects shows the low linear polarization fraction $\lesssim 1\%$ typical of unobscured Type I objects \citep{hutsemekers2017}.

Here we also find no evidence for any significant X-ray column in two observations spanning a factor of $\simeq 4$ in luminosity. In addition, the variation of the Balmer and He lines match the expectations of photoionization models \citep[e.g.,][]{korista2004}. The line widths also show a ``breathing" anti-correlation with optical luminosity compatible with a BLR emission size $R \sim L^{1/2}$ and radial velocity $v \sim R^{-1/2}$. The evolution of the emission lines quantitatively demonstrates that the BLR is responding to intrinsic changes in continuum luminosity.

Quasars are typically bluer when brighter \citep[e.g.,][]{cutrietal1985,vandenberk2004}. If the rest frame UV luminosity driving photoionization varies more than the optical that we measure, our responsivities could be overestimated. Our best measurements are for the Balmer lines: $\simeq 0.3-0.8$ (\autoref{tab:2}) compared to predicted values $\simeq 0.5-0.7$ \citep{korista2004}. It therefore appears likely that the photoionizing flux in the UV is varying in a similar fashion as that of our observed optical bands. From 2013-2013, the variability amplitude in the GALEX NUV light curve is similar to that of the quasi-simultaneous optical light curves from PS1. When observed at both high and relatively low states, the X-ray luminosity is also consistent with a constant value of $\alpha_{\rm OX} \simeq -1.2$.

\subsection{Combined constraints}

The combined observational constraints from the observed continuum and broad emission line evolution are as follows:

\begin{itemize}
    \item variations by factors of 10 within $\lesssim 1$ year and $3-6$ within 4 months in the rest-frame that are similar at all optical wavelengths observed (rest $2300-7000$\AA);
    \item broad emission line variability responding to the continuum variations both in terms of luminosity and line width;
    \item measured optical to X-ray SEDs that are typical of
      quasars with similar variability amplitudes in each band
      ($\alpha_{\rm OX} \simeq -1.2$);
    \item rapid variability across optical bands with blue leading
      red, implying a propagation speed $\gtrsim 0.5c$ ($0.1c$) in the
      2014 (2015-2018) data;
    \item an optical color that becomes slightly bluer when brighter during most of the light curve but with a rapid evolution to the red during 2018.
\end{itemize}

We next evaluate these constraints in terms of physical scenarios invoked for quasar optical (hyper)variability including changing look AGN.

\subsection{A high covering factor ``extreme" reprocessor?}

There has long been evidence for coordinated optical/UV variability in AGN \citep[e.g.,][]{cutrietal1985,clavel1991}. The implied propagation speed assuming a thin disk model is $\gtrsim 0.1c$ \citep[e.g.,][]{krolik1991,courvoisierclavel1991}, which is naturally explained by reprocessing of the central UV or X-ray emission. Typically the amplitude of these variations is $\simeq 10-20\%$, and the light curves are explained as reprocessing fluctuations superimposed on the more slowly varying local accretion luminosity. 

\citet{shappee2014} considered a geometric reprocessing model in a thin disk with intrinsic accretion luminosity $L_{\rm acc}$ irradiated by UV or soft X-ray emission from a central source with luminosity $L_C$\footnote{Here we are agnostic as to the origin of $L_C$, most likely either EUV emission from the inner radii of an inflated inner accretion disk or X-rays from the corona.} at a height $h$ above the black hole. Assuming Newtonian gravity, face-on viewing, and zero albedo, the resulting effective temperature at each radius $R$ can be written \citep[see also][]{kazanas2001,cackett2007}, 

\begin{equation}
    \sigma T_{\rm eff}^4 = \frac{3 L_{\rm acc} R_{\rm in}}{4\pi R^3} f\left(\frac{R}{R_{\rm in}}\right) + \frac{L_C(t-\tau(R))\,h}{4\pi (h^2+R^2)^{3/2}},
\end{equation}

\noindent where $f(x)=1-x^{-1/2}$ \citep{shakura1973} and 

\begin{equation}
    \tau(R) = \frac{h}{c} \left[1+\left(1+\frac{R^2}{h^2}\right)^{1/2}\right]
\end{equation}

\noindent is the geometric time delay. The time-dependent spectrum is obtained by integrating over the disk:

\begin{equation}
    \nu L_\nu = \frac{(4\pi)^2 h\nu^4}{c^2} \int_{R_{\rm in}}^\infty \frac{R dR}{\exp{(h\nu/kT_{\rm eff})}-1}.
\end{equation}

For \sdssj\, the variability amplitude is a factor of several to ten in the optical. In the model, the light curve consists of contributions both from constant (or slowly varying) local dissipation ($L_{\rm acc})$, and from the rapidly varying central source ($L_C$). To match an optical luminosity of $L_{\rm opt} \simeq 10^{44-45}\,\ergs$ requires a central luminosity $\sim 10^{45-46} \,\ergs$, about $\gtrsim 10\times$ higher than the observed optical luminosity due to geometric dilution of the flux propagating out to the optical emission radius $\simeq 50 r_g$. Instead, when observed the NUV luminosity is similar and the X-ray luminosity is a factor of few below the optical. The SED would therefore need to peak sharply at unobserved wavelengths in the far UV or very soft X-ray.

Further, the observed variability amplitude in the optical requires that the roughly constant local contribution is negligible: $L_{\rm acc} \lesssim 10^{44} \,\ergs$. This scenario requires that $\lesssim 1\%$ of the total accretion luminosity of the central source is dissipated farther out in the disk in the optical in this scenario. In most AGN, $L_{\rm UV}$ and $L_X$ are comparable to $L_{\rm opt}$. Further, the line luminosity evolution tracks that of the optical continuum with responsivities close to those expected for photoionization without any need for a large, unseen far UV/soft X-ray spectral peak.

A speculative scenario that could alleviate these problems is reprocessing in a high covering factor shell of material around the
quasar. The covering factor would need to be order unity to allow $L_{\rm opt} / L_{\rm
  UV,X} \sim 1$. Such a scenario has been considered to explain
optical tidal disruption flares \citep{guillochon2014,roth2016}. The
reprocessing layer would need to be optically thick to absorb the
incident UV or X-ray flux. It would need to be sufficiently thin not to dampen the intrinsic variability either from photon diffusion or variable light travel time to different regions. We see no evidence for such a
reprocessing layer in the X-ray spectrum (low $N_H$ consistent with being unobscured). The X-rays would therefore need to be observed along a low obscuration line of sight, e.g., near the pole. Further, a sudden change in shell radius, thickness, or geometry would still be needed to account for the spectral evolution towards the red seen in 2018.

\subsection{Rapid accretion disk evolution?}

Alternative models for quasar variability invoke mass accretion
\citep{lyubarskii1997} and/or thermal \citep{kelly2009,dexter2011}
fluctuations. Similar ideas have been proposed for changing look AGN
\citep{ross2018,noda2018,dexter2019}, where the large variations
result from either disk instabilities or large accretion rate
changes. The high cadence observations of \sdssj\ show blue leading red with a short lag \citep[$\simeq 3-6$ days,][]{homayouni2018}. At least for the variability in 2014, that behavior would seem to rule out a pure inflow scenario, where the red bands should lead. Disk instabilities are thought to drive repeating outbursts, most clearly in dwarf novae systems \citep{osaki1996}. Similar limit cycle instabilities have been proposed to explain changing look quasars \citep{ross2018,noda2018,sniegowska2019}. 

A heating front  associated with such a cycle \citep[e.g.,][]{menou1999} could drive rapid variability with blue leading red if launched from the inner edge of a geometrically thick disk. That scenario should also predict longer variability timescales for fronts launched at larger radius, and red leading blue if they propagate inwards rather than outwards. Detailed models are needed to explore whether such a scenario can produce large amplitude variations on timescales of months that are coordinated across the optical band on timescales of a few days. A potential advantage in this scenario is that dramatic reddening of the SED could result from changes to the disk structure \citep[e.g.,][]{neilsen2011}. This trend is otherwise the opposite of the prediction of optically thick disk emission, where $L \sim T_{\rm eff}^4$ as seen in the soft states of X-ray binaries \citep[e.g.,][]{gierlinski2004}. 

\subsection{Comparison to other changing look AGN}

The rapid coordinated variability between optical bands favors a reprocessing origin for the entire optical SED. At the same time, a basic reprocessing model of a central source in a thin accretion disk geometry is excluded on energetic grounds. A similar problem occurs in several other changing look AGN where $L_X \lesssim L_{\rm opt}$ \citep[][but see \citealt{noda2018}]{husemann2016,ruan2019}. 

The Seyfert galaxy NGC 2617, studied by also showed short continuum lags with blue leading red as expected for reprocessing \citet{shappee2014}. In that case, the observed $L_X$ was compatible with driving the observed variations in a thin disk geometry. Observationally, the crucial difference there is that the optical rms amplitude fluctuations were relatively small and decreased with increasing wavelength. For \sdssj, the optical varies by a factor $\simeq 10$ in all bands. As a result, a reprocessing scenario requires central UV/X-rays to drive the entire optical SED. That sets stringent limits on both the outer disk ($L_{\rm acc} < 10^{44} \,\ergs$) and central ($L_C \sim 10^{46} \,\ergs$) luminosity. Neither possibility is consistent with observations and we would need a high covering factor reprocessor (sufficiently high to intercept order unity of the central radiation) or some other mechanism to produce rapid, coordinated variability.

The observed variations are continuous over the past 10 years, with an increase by a factor $\simeq 2-3$ before the ``changing look" event and continuous large amplitude changes since. We favor the proposal that changing look quasars represent the high amplitude tail to the distribution of ordinary quasar optical variability \citep{rumbaugh2018}. The light curve is in this sense quite different from some optical changing look AGN in Seyfert galaxies. For example, the UV luminosity of Fairall 9 steadily decreased by a factor of $30$ in UV luminosity over a period of 5 years \citep{clavel1989}. Mrk 1018 remained in the high state with a relatively low optical rms for $\gtrsim 10$ years before dropping in luminosity by a factor of 10 \citep{mcelroy2016}. Mrk 590 displays a similar secular evolution while decreasing in luminosity by a factor of $100$ \citep{denney2014}. It could be that more than one mechanism is required to explain the diversity of changing look AGN. 

\section{conclusions}
\label{sec:conclusions}

We present extensive monitoring data, particularly SDSS optical spectra, from the hypervariable/changing look quasar \sdssj. We find: 

\begin{itemize}

\item All observed broad emission lines respond to the observed continuum changes. Since the BLR sees the same continuum changes that we do, they are intrinsic to the source. The line response adds to mounting evidence for intrinsic changes in accretion power as the origin of hypervariable and optical changing look AGN. The continuous variability also argues against an origin as a discrete event.

\item The observed BLR response is consistent with that predicted by photoionization models. The luminosity of the Balmer and \mgii\, lines is sub-linear while that of the He lines is nearly linear. The line widths also decrease with increasing luminosity in the fashion expected for a stratified, virialized or rotating BLR. There is no evidence for a change in intrinsic BLR structure or physical properties associated with the changing look event.

\item The continuum evolution poses a challenge to all proposed models. The rapid coordination between optical bands with blue leading red points towards a reprocessing origin, but the large variability amplitudes in all bands would require a high covering factor reprocessor (e.g., not a thin disk). This coordination and the rapid variability timescales of months are challenging for disk instability models. They may remain viable if the instability is launched at the inner edge of a geometrically thick disk.

\end{itemize}

Future studies can use time-resolved spectral evolution to probe accretion and BLR physics in a sample of hypervariable quasars out to high redshift.

\acknowledgments
JD thanks J.J. Ruan, J.H. Krolik, and R.I. Davies for helpful discussions. JD was supported by a Sofja Kovalevskaja award from the Alexander von Humboldt foundation. Y.S. acknowledges support from an Alfred P. Sloan Research Fellowship and NSF grant AST-1715579. CJG and WNB acknowledge support from NSF grant AST1517113. PH acknowledges support from NSERC grant 2017-05983.
KH acknowledges support from STFC grant ST/R000824/1. 

Funding for the Sloan Digital Sky Survey IV has been provided by the Alfred P. Sloan Foundation, the U.S. Department of Energy Office of Science, and the Participating Institutions. SDSS-IV acknowledges
support and resources from the Center for High-Performance Computing at the University of Utah. The SDSS web site is www.sdss.org.

SDSS-IV is managed by the Astrophysical Research Consortium for the 
Participating Institutions of the SDSS Collaboration including the 
Brazilian Participation Group, the Carnegie Institution for Science, 
Carnegie Mellon University, the Chilean Participation Group, the French Participation Group, Harvard-Smithsonian Center for Astrophysics, 
Instituto de Astrof\'isica de Canarias, The Johns Hopkins University, Kavli Institute for the Physics and Mathematics of the Universe (IPMU) / 
University of Tokyo, the Korean Participation Group, Lawrence Berkeley National Laboratory, 
Leibniz Institut f\"ur Astrophysik Potsdam (AIP),  
Max-Planck-Institut f\"ur Astronomie (MPIA Heidelberg), 
Max-Planck-Institut f\"ur Astrophysik (MPA Garching), 
Max-Planck-Institut f\"ur Extraterrestrische Physik (MPE), 
National Astronomical Observatories of China, New Mexico State University, 
New York University, University of Notre Dame, 
Observat\'ario Nacional / MCTI, The Ohio State University, 
Pennsylvania State University, Shanghai Astronomical Observatory, 
United Kingdom Participation Group,
Universidad Nacional Aut\'onoma de M\'exico, University of Arizona, 
University of Colorado Boulder, University of Oxford, University of Portsmouth, 
University of Utah, University of Virginia, University of Washington, University of Wisconsin, 
Vanderbilt University, and Yale University.

Based in part on observations obtained with \\MegaPrime/MegaCam, a joint project of CFHT and CEA/DAPNIA, at the Canada-France-Hawaii Telescope (CFHT) which is operated by the National Research Council (NRC) of Canada, the Institut National des Sciences de l'Univers of the Centre National de la Recherche Scientifique of France, and the University of Hawaii.
The authors wish to recognize and acknowledge the very significant
cultural role and reverence that the summit of Maunakea has always had
within the indigenous Hawaiian community.  We are most fortunate to
have the opportunity to conduct observations from this mountain.

\vspace{5mm}
\facilities{SDSS,CFHT}


\software{astropy, numpy, matplotlib}

\bibliography{sdssj}

\begin{thebibliography}{}
\expandafter\ifx\csname natexlab\endcsname\relax\def\natexlab#1{#1}\fi
\providecommand{\url}[1]{\href{#1}{#1}}

\bibitem[{{Alard}(2000)}]{alard2000}
{Alard}, C. 2000, Astronomy and Astrophysics Supplement Series, 144, 363

\bibitem[{{Alard} \& {Lupton}(1998)}]{alard1998}
{Alard}, C., \& {Lupton}, R.~H. 1998, \apj, 503, 325

\bibitem[{{Astropy Collaboration} {et~al.}(2013){Astropy Collaboration},
  {Robitaille}, {Tollerud}, {Greenfield}, {Droettboom}, {Bray}, {Aldcroft},
  {Davis}, {Ginsburg}, {Price-Whelan}, {Kerzendorf}, {Conley}, {Crighton},
  {Barbary}, {Muna}, {Ferguson}, {Grollier}, {Parikh}, {Nair}, {Unther},
  {Deil}, {Woillez}, {Conseil}, {Kramer}, {Turner}, {Singer}, {Fox}, {Weaver},
  {Zabalza}, {Edwards}, {Azalee Bostroem}, {Burke}, {Casey}, {Crawford},
  {Dencheva}, {Ely}, {Jenness}, {Labrie}, {Lim}, {Pierfederici}, {Pontzen},
  {Ptak}, {Refsdal}, {Servillat}, \& {Streicher}}]{astropy}
{Astropy Collaboration}, {Robitaille}, T.~P., {Tollerud}, E.~J., {et~al.} 2013,
  \aap, 558, A33

\bibitem[{{Aune} {et~al.}(2003){Aune}, {Boulade}, {Charlot}, {Abbon},
  {Borgeaud}, {Carton}, {Carty}, {Da Costa}, {Desforge}, {Deschamps},
  {Eppell{\'e}}, {Gallais}, {Gosset}, {Granelli}, {Gros}, {de Kat}, {Loiseau},
  {Ritou}, {Rouss{\'e}}, {Starzynski}, {Vignal}, \& {Vigroux}}]{aune2003}
{Aune}, S., {Boulade}, O., {Charlot}, X., {et~al.} 2003, in Society of
  Photo-Optical Instrumentation Engineers (SPIE) Conference Series, Vol. 4841,
  Instrument Design and Performance for Optical/Infrared Ground-based
  Telescopes, ed. M.~{Iye} \& A.~F.~M. {Moorwood}, 513--524

\bibitem[{{Baldwin}(1977)}]{baldwin1977}
{Baldwin}, J.~A. 1977, \apj, 214, 679

\bibitem[{{Blanton} {et~al.}(2017){Blanton}, {Bershady}, {Abolfathi},
  {Albareti}, {Allende Prieto}, {Almeida}, {Alonso-Garc{\'\i}a}, {Anders},
  {Anderson}, \& {Andrews}}]{blanton2017}
{Blanton}, M.~R., {Bershady}, M.~A., {Abolfathi}, B., {et~al.} 2017, \aj, 154,
  28

\bibitem[{{Bruce} {et~al.}(2017){Bruce}, {Lawrence}, {MacLeod}, {Elvis},
  {Ward}, {Collinson}, {Gezari}, {Marshall}, {Lam}, {Kotak}, {Inserra},
  {Polshaw}, {Kaiser}, {Kudritzki}, {Magnier}, \& {Waters}}]{bruce2017}
{Bruce}, A., {Lawrence}, A., {MacLeod}, C., {et~al.} 2017, \mnras, 467, 1259

\bibitem[{{Cackett} \& {Horne}(2006)}]{cackett2006}
{Cackett}, E.~M., \& {Horne}, K. 2006, \mnras, 365, 1180

\bibitem[{{Cackett} {et~al.}(2007){Cackett}, {Horne}, \&
  {Winkler}}]{cackett2007}
{Cackett}, E.~M., {Horne}, K., \& {Winkler}, H. 2007, \mnras, 380, 669

\bibitem[{{Clavel} {et~al.}(1989){Clavel}, {Wamsteker}, \&
  {Glass}}]{clavel1989}
{Clavel}, J., {Wamsteker}, W., \& {Glass}, I.~S. 1989, \apj, 337, 236

\bibitem[{{Clavel} {et~al.}(1991){Clavel}, {Reichert}, {Alloin}, {Crenshaw},
  {Kriss}, {Krolik}, {Malkan}, {Netzer}, {Peterson}, {Wamsteker}, {Altamore},
  {Baribaud}, {Barr}, {Beck}, {Binette}, {Bromage}, {Brosch}, {Diaz},
  {Filippenko}, {Fricke}, {Gaskell}, {Giommi}, {Glass}, {Gondhalekar},
  {Hackney}, {Halpern}, {Hutter}, {Joersaeter}, {Kinney}, {Kollatschny},
  {Koratkar}, {Korista}, {Laor}, {Lasota}, {Leibowitz}, {Maoz}, {Martin},
  {Mazeh}, {Meurs}, {Nair}, {O'Brien}, {Pelat}, {Perez}, {Perola}, {Ptak},
  {Rodriguez-Pascual}, {Rosenblatt}, {Sadun}, {Santos-Lleo}, {Shaw}, {Smith},
  {Stirpe}, {Stoner}, {Sun}, {Ulrich}, {van Groningen}, \&
  {Zheng}}]{clavel1991}
{Clavel}, J., {Reichert}, G.~A., {Alloin}, D., {et~al.} 1991, \apj, 366, 64

\bibitem[{{Cohen} {et~al.}(1986){Cohen}, {Rudy}, {Puetter}, {Ake}, \&
  {Foltz}}]{cohen1986}
{Cohen}, R.~D., {Rudy}, R.~J., {Puetter}, R.~C., {Ake}, T.~B., \& {Foltz},
  C.~B. 1986, \apj, 311, 135

\bibitem[{{Collinson} {et~al.}(2018){Collinson}, {Ward}, {Lawrence}, {Bruce},
  {MacLeod}, {Elvis}, {Gezari}, {Marshall}, \& {Done}}]{collinson2018}
{Collinson}, J.~S., {Ward}, M.~J., {Lawrence}, A., {et~al.} 2018, \mnras, 474,
  3565

\bibitem[{{Courvoisier} \& {Clavel}(1991)}]{courvoisierclavel1991}
{Courvoisier}, T., \& {Clavel}, J. 1991, \aap, 248, 389

\bibitem[{{Cutri} {et~al.}(1985){Cutri}, {Wisniewski}, {Rieke}, \&
  {Lebofsky}}]{cutrietal1985}
{Cutri}, R.~M., {Wisniewski}, W.~Z., {Rieke}, G.~H., \& {Lebofsky}, M.~J. 1985,
  \apj, 296, 423

\bibitem[{{Denney} {et~al.}(2014){Denney}, {De Rosa}, {Croxall}, {Gupta},
  {Bentz}, {Fausnaugh}, {Grier}, {Martini}, {Mathur}, {Peterson}, {Pogge}, \&
  {Shappee}}]{denney2014}
{Denney}, K.~D., {De Rosa}, G., {Croxall}, K., {et~al.} 2014, \apj, 796, 134

\bibitem[{{Dexter} \& {Agol}(2011)}]{dexter2011}
{Dexter}, J., \& {Agol}, E. 2011, \apjl, 727, L24

\bibitem[{{Dexter} \& {Begelman}(2019)}]{dexter2019}
{Dexter}, J., \& {Begelman}, M.~C. 2019, \mnras, 483, L17

\bibitem[{{Edelson} {et~al.}(2019){Edelson}, {Gelbord}, {Cackett}, {Peterson},
  {Horne}, {Barth}, {Starkey}, {Bentz}, {Brandt}, {Goad}, {Joner}, {Korista},
  {Netzer}, {Page}, {Uttley}, {Vaughan}, {Breeveld}, {Cenko}, {Done}, {Evans},
  {Fausnaugh}, {Ferland}, {Gonzalez-Buitrago}, {Gropp}, {Grupe}, {Kaastra},
  {Kennea}, {Kriss}, {Mathur}, {Mehdipour}, {Mudd}, {Nousek}, {Schmidt},
  {Vestergaard}, \& {Villforth}}]{edelson2019}
{Edelson}, R., {Gelbord}, J., {Cackett}, E., {et~al.} 2019, \apj, 870, 123

\bibitem[{{Fausnaugh} {et~al.}(2016){Fausnaugh}, {Denney}, {Barth}, {Bentz},
  {Bottorff}, {Carini}, {Croxall}, {De Rosa}, {Goad}, {Horne}, {Joner},
  {Kaspi}, {Kim}, {Klimanov}, {Kochanek}, {Leonard}, {Netzer}, {Peterson},
  {Schn{\"u}lle}, {Sergeev}, {Vestergaard}, {Zheng}, {Zu}, {Anderson},
  {Ar{\'e}valo}, {Bazhaw}, {Borman}, {Boroson}, {Brandt}, {Breeveld}, {Brewer},
  {Cackett}, {Crenshaw}, {Dalla Bont{\`a}}, {De Lorenzo-C{\'a}ceres},
  {Dietrich}, {Edelson}, {Efimova}, {Ely}, {Evans}, {Filippenko}, {Flatland},
  {Gehrels}, {Geier}, {Gelbord}, {Gonzalez}, {Gorjian}, {Grier}, {Grupe},
  {Hall}, {Hicks}, {Horenstein}, {Hutchison}, {Im}, {Jensen}, {Jones},
  {Kaastra}, {Kelly}, {Kennea}, {Kim}, {Korista}, {Kriss}, {Lee}, {Lira},
  {MacInnis}, {Manne-Nicholas}, {Mathur}, {McHardy}, {Montouri}, {Musso},
  {Nazarov}, {Norris}, {Nousek}, {Okhmat}, {Pancoast}, {Papadakis}, {Parks},
  {Pei}, {Pogge}, {Pott}, {Rafter}, {Rix}, {Saylor}, {Schimoia}, {Siegel},
  {Spencer}, {Starkey}, {Sung}, {Teems}, {Treu}, {Turner}, {Uttley},
  {Villforth}, {Weiss}, {Woo}, {Yan}, \& {Young}}]{fausnaugh2016}
{Fausnaugh}, M.~M., {Denney}, K.~D., {Barth}, A.~J., {et~al.} 2016, \apj, 821,
  56

\bibitem[{{Fukugita} {et~al.}(1996){Fukugita}, {Ichikawa}, {Gunn}, {Doi},
  {Shimasaku}, \& {Schneider}}]{fukugita1996}
{Fukugita}, M., {Ichikawa}, T., {Gunn}, J.~E., {et~al.} 1996, \aj, 111, 1748

\bibitem[{{Gezari} {et~al.}(2013){Gezari}, {Martin}, {Forster}, {Neill},
  {Huber}, {Heckman}, {Bianchi}, {Morrissey}, {Neff}, {Seibert},
  {Schiminovich}, {Wyder}, {Burgett}, {Chambers}, {Kaiser}, {Magnier}, {Price},
  \& {Tonry}}]{gezari2013}
{Gezari}, S., {Martin}, D.~C., {Forster}, K., {et~al.} 2013, \apj, 766, 60

\bibitem[{{Gezari} {et~al.}(2017){Gezari}, {Hung}, {Cenko}, {Blagorodnova},
  {Yan}, {Kulkarni}, {Mooley}, {Kong}, {Cantwell}, {Yu}, {Cao}, {Fremling},
  {Neill}, {Ngeow}, {Nugent}, \& {Wozniak}}]{gezari2017}
{Gezari}, S., {Hung}, T., {Cenko}, S.~B., {et~al.} 2017, \apj, 835, 144

\bibitem[{{Gierli{\'n}ski} \& {Done}(2004)}]{gierlinski2004}
{Gierli{\'n}ski}, M., \& {Done}, C. 2004, \mnras, 347, 885

\bibitem[{{Grier} {et~al.}(2017){Grier}, {Trump}, {Shen}, {Horne}, {Kinemuchi},
  {McGreer}, {Starkey}, {Brandt}, {Hall}, {Kochanek}, {Chen}, {Denney},
  {Greene}, {Ho}, {Homayouni}, {I-Hsiu Li}, {Pei}, {Peterson}, {Petitjean},
  {Schneider}, {Sun}, {AlSayyad}, {Bizyaev}, {Brinkmann}, {Brownstein},
  {Bundy}, {Dawson}, {Eftekharzadeh}, {Fernandez-Trincado}, {Gao},
  {Hutchinson}, {Jia}, {Jiang}, {Oravetz}, {Pan}, {Paris}, {Ponder}, {Peters},
  {Rogerson}, {Simmons}, {Smith}, \& {Wang}}]{grier2017}
{Grier}, C.~J., {Trump}, J.~R., {Shen}, Y., {et~al.} 2017, \apj, 851, 21

\bibitem[{{Grier} {et~al.}(2019){Grier}, {Shen}, {Horne}, {Brandt}, {Trump},
  {Kinemuchi}, {Schneider}, {Homayouni}, {McGreer}, {Peterson}, {Bizyaev},
  {Chen}, {Dawson}, {Eftekharzadeh}, {Kneib}, {Nie}, {Oravetz}, {Oravetz},
  {Pan}, {Petitjean}, {Vivek}, \& {Zou}}]{grier2019}
{Grier}, C.~J., {Shen}, Y., {Horne}, K., {et~al.} 2019, arXiv e-prints,
  arXiv:1904.03199

\bibitem[{{Guillochon} {et~al.}(2014){Guillochon}, {Manukian}, \&
  {Ramirez-Ruiz}}]{guillochon2014}
{Guillochon}, J., {Manukian}, H., \& {Ramirez-Ruiz}, E. 2014, \apj, 783, 23

\bibitem[{{Gunn} {et~al.}(2006){Gunn}, {Siegmund}, {Mannery}, {Owen}, {Hull},
  {Leger}, {Carey}, {Knapp}, {York}, \& {Boroski}}]{gunn2006}
{Gunn}, J.~E., {Siegmund}, W.~A., {Mannery}, E.~J., {et~al.} 2006, \aj, 131,
  2332

\bibitem[{{Hinshaw} {et~al.}(2013){Hinshaw}, {Larson}, {Komatsu}, {Spergel},
  {Bennett}, {Dunkley}, {Nolta}, {Halpern}, {Hill}, {Odegard}, {Page}, {Smith},
  {Weiland}, {Gold}, {Jarosik}, {Kogut}, {Limon}, {Meyer}, {Tucker}, {Wollack},
  \& {Wright}}]{hinshaw2013}
{Hinshaw}, G., {Larson}, D., {Komatsu}, E., {et~al.} 2013, \apjs, 208, 19

\bibitem[{{Homayouni} {et~al.}(2018){Homayouni}, {Trump}, {Grier}, {Shen},
  {Starkey}, {Brandt}, {Hall}, {Horne}, {Kinemuchi}, {I-Hsiu Li}, {McGreer},
  {Sun}, {Ho}, \& {Schneider}}]{homayouni2018}
{Homayouni}, Y., {Trump}, J.~R., {Grier}, C.~J., {et~al.} 2018, arXiv e-prints,
  arXiv:1806.08360

\bibitem[{{Husemann} {et~al.}(2016){Husemann}, {Urrutia}, {Tremblay}, {Krumpe},
  {Dexter}, {Busch}, {Combes}, {Croom}, {Davis}, {Eckart}, {McElroy},
  {Perez-Torres}, {Powell}, \& {Scharw{\"a}chter}}]{husemann2016}
{Husemann}, B., {Urrutia}, T., {Tremblay}, G.~R., {et~al.} 2016, \aap, 593, L9

\bibitem[{{Hutsem{\'e}kers} {et~al.}(2017){Hutsem{\'e}kers}, {Ag{\'{\i}}s
  Gonz{\'a}lez}, {Sluse}, {Ramos Almeida}, \& {Acosta
  Pulido}}]{hutsemekers2017}
{Hutsem{\'e}kers}, D., {Ag{\'{\i}}s Gonz{\'a}lez}, B., {Sluse}, D., {Ramos
  Almeida}, C., \& {Acosta Pulido}, J.-A. 2017, \aap, 604, L3

\bibitem[{{Kazanas} \& {Nayakshin}(2001)}]{kazanas2001}
{Kazanas}, D., \& {Nayakshin}, S. 2001, \apj, 550, 655

\bibitem[{{Kelly} {et~al.}(2009){Kelly}, {Bechtold}, \&
  {Siemiginowska}}]{kelly2009}
{Kelly}, B.~C., {Bechtold}, J., \& {Siemiginowska}, A. 2009, \apj, 698, 895

\bibitem[{{Kim} {et~al.}(2018){Kim}, {Yoon}, \& {Evans}}]{kim2018}
{Kim}, D.~C., {Yoon}, I., \& {Evans}, A.~S. 2018, \apj, 861, 51

\bibitem[{{Koratkar} \& {Gaskell}(1991)}]{koratkar1991}
{Koratkar}, A.~P., \& {Gaskell}, C.~M. 1991, \apjl, 370, L61

\bibitem[{{Korista} \& {Goad}(2004)}]{korista2004}
{Korista}, K.~T., \& {Goad}, M.~R. 2004, \apj, 606, 749

\bibitem[{{Koz{\l}owski} {et~al.}(2010){Koz{\l}owski}, {Kochanek}, {Udalski},
  {Wyrzykowski}, {Soszy{\'n}ski}, {Szyma{\'n}ski}, {Kubiak}, {Pietrzy{\'n}ski},
  {Szewczyk}, {Ulaczyk}, {Poleski}, \& {The OGLE
  Collaboration}}]{kozlowski2010}
{Koz{\l}owski}, S., {Kochanek}, C.~S., {Udalski}, A., {et~al.} 2010, \apj, 708,
  927

\bibitem[{{Krolik} {et~al.}(1991){Krolik}, {Horne}, {Kallman}, {Malkan},
  {Edelson}, \& {Kriss}}]{krolik1991}
{Krolik}, J.~H., {Horne}, K., {Kallman}, T.~R., {et~al.} 1991, \apj, 371, 541

\bibitem[{{LaMassa} {et~al.}(2015){LaMassa}, {Cales}, {Moran}, {Myers},
  {Richards}, {Eracleous}, {Heckman}, {Gallo}, \& {Urry}}]{lamassa2015}
{LaMassa}, S.~M., {Cales}, S., {Moran}, E.~C., {et~al.} 2015, \apj, 800, 144

\bibitem[{{Lawrence}(2018)}]{lawrence2018}
{Lawrence}, A. 2018, Nature Astronomy, 2, 102

\bibitem[{{Lusso} {et~al.}(2010){Lusso}, {Comastri}, {Vignali}, {Zamorani},
  {Brusa}, {Gilli}, {Iwasawa}, {Salvato}, {Civano}, {Elvis}, {Merloni},
  {Bongiorno}, {Trump}, {Koekemoer}, {Schinnerer}, {Le Floc'h}, {Cappelluti},
  {Jahnke}, {Sargent}, {Silverman}, {Mainieri}, {Fiore}, {Bolzonella}, {Le
  F{\`e}vre}, {Garilli}, {Iovino}, {Kneib}, {Lamareille}, {Lilly}, {Mignoli},
  {Scodeggio}, \& {Vergani}}]{lusso2010}
{Lusso}, E., {Comastri}, A., {Vignali}, C., {et~al.} 2010, \aap, 512, A34

\bibitem[{{Lyubarskii}(1997)}]{lyubarskii1997}
{Lyubarskii}, Y.~E. 1997, \mnras, 292, 679

\bibitem[{{MacLeod} {et~al.}(2010){MacLeod}, {Ivezi{\'c}}, {Kochanek},
  {Koz{\l}owski}, {Kelly}, {Bullock}, {Kimball}, {Sesar}, {Westman}, {Brooks},
  {Gibson}, {Becker}, \& {de Vries}}]{macleod2010}
{MacLeod}, C.~L., {Ivezi{\'c}}, {\v Z}., {Kochanek}, C.~S., {et~al.} 2010,
  \apj, 721, 1014

\bibitem[{{MacLeod} {et~al.}(2016){MacLeod}, {Ross}, {Lawrence}, {Goad},
  {Horne}, {Burgett}, {Chambers}, {Flewelling}, {Hodapp}, {Kaiser}, {Magnier},
  {Wainscoat}, \& {Waters}}]{macleod2016}
{MacLeod}, C.~L., {Ross}, N.~P., {Lawrence}, A., {et~al.} 2016, \mnras, 457,
  389

\bibitem[{{MacLeod} {et~al.}(2018){MacLeod}, {Green}, {Anderson}, {Eracleous},
  {Ruan}, {Runnoe}, {Nielsen Brandt}, {Badenes}, {Greene}, {Morganson},
  {Schmidt}, {Schwope}, {Shen}, {Amaro}, {Lebleu}, {Filiz Ak}, {Grier},
  {Hoover}, {McGraw}, {Dawson}, {Hall}, {Hawley}, {Mariappan}, {Myers},
  {P{\^a}ris}, {Schneider}, {Stassun}, {Bershady}, {Blanton}, {Seo}, {Tinker},
  {Fern{\'a}ndez-Trincado}, {Chambers}, {Kaiser}, {Kudritzki}, {Magnier},
  {Metcalfe}, \& {Waters}}]{macleod2018}
{MacLeod}, C.~L., {Green}, P.~J., {Anderson}, S.~F., {et~al.} 2018, \aj, 155, 6

\bibitem[{{Magnier} {et~al.}(2013){Magnier}, {Schlafly}, {Finkbeiner}, {Juric},
  {Tonry}, {Burgett}, {Chambers}, {Flewelling}, {Kaiser}, \&
  {Kudritzki}}]{magnier2013}
{Magnier}, E.~A., {Schlafly}, E., {Finkbeiner}, D., {et~al.} 2013, \apjs, 205,
  20

\bibitem[{{Matt} {et~al.}(2003){Matt}, {Guainazzi}, \& {Maiolino}}]{matt2003}
{Matt}, G., {Guainazzi}, M., \& {Maiolino}, R. 2003, \mnras, 342, 422

\bibitem[{{McElroy} {et~al.}(2016){McElroy}, {Husemann}, {Croom}, {Davis},
  {Bennert}, {Busch}, {Combes}, {Eckart}, {Perez-Torres}, {Powell},
  {Scharw{\"a}chter}, {Tremblay}, \& {Urrutia}}]{mcelroy2016}
{McElroy}, R.~E., {Husemann}, B., {Croom}, S.~M., {et~al.} 2016, \aap, 593, L8

\bibitem[{{McHardy} {et~al.}(2014){McHardy}, {Cameron}, {Dwelly}, {Connolly},
  {Lira}, {Emmanoulopoulos}, {Gelbord}, {Breedt}, {Arevalo}, \&
  {Uttley}}]{mchardy2014}
{McHardy}, I.~M., {Cameron}, D.~T., {Dwelly}, T., {et~al.} 2014, \mnras, 444,
  1469

\bibitem[{{Menou} {et~al.}(1999){Menou}, {Hameury}, \& {Stehle}}]{menou1999}
{Menou}, K., {Hameury}, J.-M., \& {Stehle}, R. 1999, \mnras, 305, 79

\bibitem[{{Merloni} {et~al.}(2015){Merloni}, {Dwelly}, {Salvato},
  {Georgakakis}, {Greiner}, {Krumpe}, {Nandra}, {Ponti}, \&
  {Rau}}]{merloni2015}
{Merloni}, A., {Dwelly}, T., {Salvato}, M., {et~al.} 2015, \mnras, 452, 69

\bibitem[{{Neilsen} {et~al.}(2011){Neilsen}, {Remillard}, \&
  {Lee}}]{neilsen2011}
{Neilsen}, J., {Remillard}, R.~A., \& {Lee}, J.~C. 2011, \apj, 737, 69

\bibitem[{{Noda} \& {Done}(2018)}]{noda2018}
{Noda}, H., \& {Done}, C. 2018, ArXiv e-prints, arXiv:1805.07873

\bibitem[{{Osaki}(1996)}]{osaki1996}
{Osaki}, Y. 1996, \pasp, 108, 39

\bibitem[{{Peterson} {et~al.}(1991){Peterson}, {Balonek}, {Barker}, {Bechtold},
  {Bertram}, {Bochkarev}, {Bolte}, {Bond}, {Boroson}, {Carini}, {Carone},
  {Christensen}, {Clements}, {Cochran}, {Cohen}, {Crampton}, {Dietrich},
  {Elvis}, {Ferguson}, {Filippenko}, {Fricke}, {Gaskell}, {Halpern}, {Huchra},
  {Hutchings}, {Kollatschny}, {Koratkar}, {Korista}, {Krolik}, {Lame}, {Laor},
  {Leacock}, {MacAlpine}, {Malkan}, {Maoz}, {Miller}, {Morris}, {Netzer},
  {Oliveira}, {Penfold}, {Penston}, {Perez}, {Pogge}, {Richmond}, {Romanishin},
  {Rosenblatt}, {Saddlemyer}, {Sadun}, {Sawyer}, {Shields}, {Shapovalova},
  {Smith}, {Smith}, {Smith}, {Sun}, {Thiele}, {Turner}, {Veilleux}, {Wagner},
  {Weymann}, {Wilkes}, {Wills}, {Wills}, \& {Younger}}]{peterson1991}
{Peterson}, B.~M., {Balonek}, T.~J., {Barker}, E.~S., {et~al.} 1991, \apj, 368,
  119

\bibitem[{{Peterson} {et~al.}(2002){Peterson}, {Berlind}, {Bertram},
  {Bischoff}, {Bochkarev}, {Borisov}, {Burenkov}, {Calkins}, {Carrasco},
  {Chavushyan}, {Chornock}, {Dietrich}, {Doroshenko}, {Ezhkova}, {Filippenko},
  {Gilbert}, {Huchra}, {Kollatschny}, {Leonard}, {Li}, {Lyuty}, {Malkov},
  {Matheson}, {Merkulova}, {Mikhailov}, {Modjaz}, {Onken}, {Pogge}, {Pronik},
  {Qian}, {Romano}, {Sergeev}, {Sergeeva}, {Shapovalova}, {Spiridonova}, {Tao},
  {Tokarz}, {Valdes}, {Vlasiuk}, {Wagner}, \& {Wilkes}}]{peterson2002}
{Peterson}, B.~M., {Berlind}, P., {Bertram}, R., {et~al.} 2002, \apj, 581, 197

\bibitem[{{Pringle}(1981)}]{pringle1981}
{Pringle}, J.~E. 1981, \araa, 19, 137

\bibitem[{{Risaliti} {et~al.}(2002){Risaliti}, {Elvis}, \&
  {Nicastro}}]{risaliti2002}
{Risaliti}, G., {Elvis}, M., \& {Nicastro}, F. 2002, \apj, 571, 234

\bibitem[{{Ross} {et~al.}(2018){Ross}, {Ford}, {Graham}, {McKernan}, {Stern},
  {Meisner}, {Asse}, {Dey}, {Drake}, \& {Jun}}]{ross2018}
{Ross}, N.~P., {Ford}, K.~E.~S., {Graham}, M., {et~al.} 2018, ArXiv e-prints,
  arXiv:1805.06921

\bibitem[{{Roth} {et~al.}(2016){Roth}, {Kasen}, {Guillochon}, \&
  {Ramirez-Ruiz}}]{roth2016}
{Roth}, N., {Kasen}, D., {Guillochon}, J., \& {Ramirez-Ruiz}, E. 2016, \apj,
  827, 3

\bibitem[{{Ruan} {et~al.}(2019){Ruan}, {Anderson}, {Eracleous}, {Green},
  {Haggard}, {MacLeod}, {Runnoe}, \& {Sobolewska}}]{ruan2019}
{Ruan}, J.~J., {Anderson}, S.~F., {Eracleous}, M., {et~al.} 2019, arXiv
  e-prints, arXiv:1903.02553

\bibitem[{{Ruan} {et~al.}(2016){Ruan}, {Anderson}, {Cales}, {Eracleous},
  {Green}, {Morganson}, {Runnoe}, {Shen}, {Wilkinson}, {Blanton}, {Dwelly},
  {Georgakakis}, {Greene}, {LaMassa}, {Merloni}, \& {Schneider}}]{ruan2016}
{Ruan}, J.~J., {Anderson}, S.~F., {Cales}, S.~L., {et~al.} 2016, \apj, 826, 188

\bibitem[{{Rumbaugh} {et~al.}(2018){Rumbaugh}, {Shen}, {Morganson}, {Liu},
  {Banerji}, {McMahon}, {Abdalla}, {Benoit-L{\'e}vy}, {Bertin}, {Brooks},
  {Buckley-Geer}, {Capozzi}, {Carnero Rosell}, {Carrasco Kind}, {Carretero},
  {Cunha}, {D'Andrea}, {da Costa}, {DePoy}, {Desai}, {Doel}, {Frieman},
  {Garc{\'{\i}}a-Bellido}, {Gruen}, {Gruendl}, {Gschwend}, {Gutierrez},
  {Honscheid}, {James}, {Kuehn}, {Kuhlmann}, {Kuropatkin}, {Lima}, {Maia},
  {Marshall}, {Martini}, {Menanteau}, {Plazas}, {Reil}, {Roodman}, {Sanchez},
  {Scarpine}, {Schindler}, {Schubnell}, {Sheldon}, {Smith}, {Soares-Santos},
  {Sobreira}, {Suchyta}, {Swanson}, {Walker}, {Wester}, \& {(DES
  Collaboration}}]{rumbaugh2018}
{Rumbaugh}, N., {Shen}, Y., {Morganson}, E., {et~al.} 2018, \apj, 854, 160

\bibitem[{{Runnoe} {et~al.}(2016){Runnoe}, {Cales}, {Ruan}, {Eracleous},
  {Anderson}, {Shen}, {Green}, {Morganson}, {LaMassa}, {Greene}, {Dwelly},
  {Schneider}, {Merloni}, {Georgakakis}, \& {Roman-Lopes}}]{runnoe2016}
{Runnoe}, J.~C., {Cales}, S., {Ruan}, J.~J., {et~al.} 2016, \mnras, 455, 1691

\bibitem[{{Schlafly} {et~al.}(2012){Schlafly}, {Finkbeiner}, {Juri{\'c}},
  {Magnier}, {Burgett}, {Chambers}, {Grav}, {Hodapp}, {Kaiser}, \&
  {Kudritzki}}]{schlafly2012}
{Schlafly}, E.~F., {Finkbeiner}, D.~P., {Juri{\'c}}, M., {et~al.} 2012, \apj,
  756, 158

\bibitem[{{Sergeev} {et~al.}(2005){Sergeev}, {Doroshenko}, {Golubinskiy},
  {Merkulova}, \& {Sergeeva}}]{sergeev2005}
{Sergeev}, S.~G., {Doroshenko}, V.~T., {Golubinskiy}, Y.~V., {Merkulova},
  N.~I., \& {Sergeeva}, E.~A. 2005, \apj, 622, 129

\bibitem[{{Shakura} \& {Sunyaev}(1973)}]{shakura1973}
{Shakura}, N.~I., \& {Sunyaev}, R.~A. 1973, \aap, 24, 337

\bibitem[{{Shappee} {et~al.}(2014){Shappee}, {Prieto}, {Grupe}, {Kochanek},
  {Stanek}, {De Rosa}, {Mathur}, {Zu}, {Peterson}, {Pogge}, {Komossa}, {Im},
  {Jencson}, {Holoien}, {Basu}, {Beacom}, {Szczygie{\l}}, {Brimacombe},
  {Adams}, {Campillay}, {Choi}, {Contreras}, {Dietrich}, {Dubberley},
  {Elphick}, {Foale}, {Giustini}, {Gonzalez}, {Hawkins}, {Howell}, {Hsiao},
  {Koss}, {Leighly}, {Morrell}, {Mudd}, {Mullins}, {Nugent}, {Parrent},
  {Phillips}, {Pojmanski}, {Rosing}, {Ross}, {Sand}, {Terndrup}, {Valenti},
  {Walker}, \& {Yoon}}]{shappee2014}
{Shappee}, B.~J., {Prieto}, J.~L., {Grupe}, D., {et~al.} 2014, \apj, 788, 48

\bibitem[{{Shen}(2013)}]{shen2013}
{Shen}, Y. 2013, Bulletin of the Astronomical Society of India, 41, 61

\bibitem[{{Shen} {et~al.}(2015{\natexlab{a}}){Shen}, {Brandt}, {Dawson},
  {Hall}, {McGreer}, {Anderson}, {Chen}, {Denney}, {Eftekharzadeh}, {Fan},
  {Gao}, {Green}, {Greene}, {Ho}, {Horne}, {Jiang}, {Kelly}, {Kinemuchi},
  {Kochanek}, {P{\^a}ris}, {Peters}, {Peterson}, {Petitjean}, {Ponder},
  {Richards}, {Schneider}, {Seth}, {Smith}, {Strauss}, {Tao}, {Trump},
  {Wood-Vasey}, {Zu}, {Eisenstein}, {Pan}, {Bizyaev}, {Malanushenko},
  {Malanushenko}, \& {Oravetz}}]{shen2015}
{Shen}, Y., {Brandt}, W.~N., {Dawson}, K.~S., {et~al.} 2015{\natexlab{a}},
  \apjs, 216, 4

\bibitem[{{Shen} {et~al.}(2015{\natexlab{b}}){Shen}, {Greene}, {Ho}, {Brandt},
  {Denney}, {Horne}, {Jiang}, {Kochanek}, {McGreer}, {Merloni}, {Peterson},
  {Petitjean}, {Schneider}, {Schulze}, {Strauss}, {Tao}, {Trump}, {Pan}, \&
  {Bizyaev}}]{shen2015msigma}
{Shen}, Y., {Greene}, J.~E., {Ho}, L.~C., {et~al.} 2015{\natexlab{b}}, \apj,
  805, 96

\bibitem[{{Shen} {et~al.}(2016){Shen}, {Horne}, {Grier}, {Peterson}, {Denney},
  {Trump}, {Sun}, {Brandt}, {Kochanek}, {Dawson}, {Green}, {Greene}, {Hall},
  {Ho}, {Jiang}, {Kinemuchi}, {McGreer}, {Petitjean}, {Richards}, {Schneider},
  {Strauss}, {Tao}, {Wood-Vasey}, {Zu}, {Pan}, {Bizyaev}, {Ge}, {Oravetz}, \&
  {Simmons}}]{shen2016}
{Shen}, Y., {Horne}, K., {Grier}, C.~J., {et~al.} 2016, \apj, 818, 30

\bibitem[{{Shen} {et~al.}(2018){Shen}, {Hall}, {Horne}, {Zhu}, {McGreer},
  {Simm}, {Trump}, {Kinemuchi}, {Brandt}, {Green}, {Grier}, {Guo}, {Ho},
  {Homayouni}, {Jiang}, {I-Hsiu Li}, {Morganson}, {Petitjean}, {Richards},
  {Schneider}, {Starkey}, {Wang}, {Chambers}, {Kaiser}, {Kudritzki}, {Magnier},
  \& {Waters}}]{shen2018}
{Shen}, Y., {Hall}, P.~B., {Horne}, K., {et~al.} 2018, arXiv e-prints,
  arXiv:1810.01447

\bibitem[{{Shen} {et~al.}(2019){Shen}, {Hall}, {Horne}, {Zhu}, {McGreer},
  {Simm}, {Trump}, {Kinemuchi}, {Brandt}, \& {Green}}]{shen2019}
---. 2019, \apjs, 241, 34

\bibitem[{{Sheng} {et~al.}(2017){Sheng}, {Wang}, {Jiang}, {Yang}, {Yan}, {Dou},
  \& {Peng}}]{sheng2017}
{Sheng}, Z., {Wang}, T., {Jiang}, N., {et~al.} 2017, \apjl, 846, L7

\bibitem[{{Shull} {et~al.}(2012){Shull}, {Stevans}, \& {Danforth}}]{shull2012}
{Shull}, J.~M., {Stevans}, M., \& {Danforth}, C.~W. 2012, \apj, 752, 162

\bibitem[{{Smee} {et~al.}(2013){Smee}, {Gunn}, {Uomoto}, {Roe}, {Schlegel},
  {Rockosi}, {Carr}, {Leger}, {Dawson}, \& {Olmstead}}]{smee2013}
{Smee}, S.~A., {Gunn}, J.~E., {Uomoto}, A., {et~al.} 2013, \aj, 146, 32

\bibitem[{{{\'S}niegowska} \& {Czerny}(2019)}]{sniegowska2019}
{{\'S}niegowska}, M., \& {Czerny}, B. 2019, arXiv e-prints, arXiv:1904.06767

\bibitem[{{Starkey} {et~al.}(2016){Starkey}, {Horne}, \&
  {Villforth}}]{starkey2016}
{Starkey}, D.~A., {Horne}, K., \& {Villforth}, C. 2016, \mnras, 456, 1960

\bibitem[{{Steffen} {et~al.}(2006){Steffen}, {Strateva}, {Brandt}, {Alexand
  er}, {Koekemoer}, {Lehmer}, {Schneider}, \& {Vignali}}]{steffen2006}
{Steffen}, A.~T., {Strateva}, I., {Brandt}, W.~N., {et~al.} 2006, \aj, 131,
  2826

\bibitem[{{Stern} {et~al.}(2018){Stern}, {McKernan}, {Graham}, {Ford}, {Ross},
  {Meisner}, {Assef}, {Balokovi{\'c}}, {Brightman}, {Dey}, {Drake},
  {Djorgovski}, {Eisenhardt}, \& {Jun}}]{stern2018}
{Stern}, D., {McKernan}, B., {Graham}, M.~J., {et~al.} 2018, ArXiv e-prints,
  arXiv:1805.06920

\bibitem[{{Storchi-Bergmann} {et~al.}(1995){Storchi-Bergmann}, {Eracleous},
  {Livio}, {Wilson}, {Filippenko}, \& {Halpern}}]{storchibergmann1995}
{Storchi-Bergmann}, T., {Eracleous}, M., {Livio}, M., {et~al.} 1995, \apj, 443,
  617

\bibitem[{{Sun} {et~al.}(2015){Sun}, {Trump}, {Shen}, {Brand t}, {Dawson},
  {Denney}, {Hall}, {Ho}, {Horne}, {Jiang}, {Richards}, {Schneider}, {Bizyaev},
  {Kinemuchi}, {Oravetz}, {Pan}, \& {Simmons}}]{sun2015}
{Sun}, M., {Trump}, J.~R., {Shen}, Y., {et~al.} 2015, \apj, 811, 42

\bibitem[{{Tohline} \& {Osterbrock}(1976)}]{tohlineosterbrock1976}
{Tohline}, J.~E., \& {Osterbrock}, D.~E. 1976, \apjl, 210, L117

\bibitem[{{Tonry} {et~al.}(2012){Tonry}, {Stubbs}, {Kilic}, {Flewelling},
  {Deacon}, {Chornock}, {Berger}, {Burgett}, {Chambers}, {Kaiser}, {Kudritzki},
  {Hodapp}, {Magnier}, {Morgan}, {Price}, \& {Wainscoat}}]{tonry2012}
{Tonry}, J.~L., {Stubbs}, C.~W., {Kilic}, M., {et~al.} 2012, \apj, 745, 42

\bibitem[{{Trakhtenbrot} {et~al.}(2019{\natexlab{a}}){Trakhtenbrot}, {Arcavi},
  {MacLeod}, {Ricci}, {Kara}, {Graham}, {Stern}, {Harrison}, {Burke},
  {Hiramatsu}, {Hosseinzadeh}, {Howell}, {Smartt}, {Rest}, {Prieto}, {Shappee},
  {Holoien}, {Bersier}, {Filippenko}, {Brink}, {Zheng}, {Li}, {Remillard}, \&
  {Loewenstein}}]{trakhtenbrot2019}
{Trakhtenbrot}, B., {Arcavi}, I., {MacLeod}, C.~L., {et~al.}
  2019{\natexlab{a}}, arXiv e-prints, arXiv:1903.11084

\bibitem[{{Trakhtenbrot} {et~al.}(2019{\natexlab{b}}){Trakhtenbrot}, {Arcavi},
  {Ricci}, {Tacchella}, {Stern}, {Netzer}, {Jonker}, {Horesh},
  {Mej{\'\i}a-Restrepo}, \& {Hosseinzadeh}}]{trakhtenbrot2019bowen}
{Trakhtenbrot}, B., {Arcavi}, I., {Ricci}, C., {et~al.} 2019{\natexlab{b}},
  Nature Astronomy, 3, 242

\bibitem[{{Vanden Berk} {et~al.}(2004){Vanden Berk}, {Wilhite}, {Kron},
  {Anderson}, {Brunner}, {Hall}, {Ivezi{\'c}}, {Richards}, {Schneider}, {York},
  {Brinkmann}, {Lamb}, {Nichol}, \& {Schlegel}}]{vandenberk2004}
{Vanden Berk}, D.~E., {Wilhite}, B.~C., {Kron}, R.~G., {et~al.} 2004, \apj,
  601, 692

\bibitem[{{Wang} {et~al.}(2018){Wang}, {Xu}, \& {Wei}}]{wang2018}
{Wang}, J., {Xu}, D.~W., \& {Wei}, J.~Y. 2018, \apj, 858, 49

\bibitem[{{Wang} {et~al.}(2019){Wang}, {Shen}, {Jiang}, {Horne}, {Brandt},
  {Grier}, {Ho}, {Homayouni}, {I-Hsiu Li}, {Schneider}, \& {Trump}}]{wang2019}
{Wang}, S., {Shen}, Y., {Jiang}, L., {et~al.} 2019, arXiv e-prints,
  arXiv:1903.10015

\bibitem[{{Williams} {et~al.}(2004){Williams}, {Olszewski}, {Lesser}, \&
  {Burge}}]{williams2004}
{Williams}, G.~G., {Olszewski}, E., {Lesser}, M.~P., \& {Burge}, J.~H. 2004, in
  Society of Photo-Optical Instrumentation Engineers (SPIE) Conference Series,
  Vol. 5492, Ground-based Instrumentation for Astronomy, ed. A.~F.~M.
  {Moorwood} \& M.~{Iye}, 787--798

\bibitem[{{Yang} {et~al.}(2016){Yang}, {Brandt}, {Luo}, {Xue}, {Bauer}, {Sun},
  {Kim}, {Schulze}, {Zheng}, {Paolillo}, {Shemmer}, {Liu}, {Schneider},
  {Vignali}, {Vito}, \& {Wang}}]{yang2016}
{Yang}, G., {Brandt}, W.~N., {Luo}, B., {et~al.} 2016, \apj, 831, 145

\bibitem[{{Yang} {et~al.}(2018){Yang}, {Wu}, {Fan}, {Jiang}, {McGreer},
  {Shangguan}, {Yao}, {Wang}, {Joshi}, {Green}, {Wang}, {Feng}, {Fu}, {Yang},
  \& {Liu}}]{yang2018}
{Yang}, Q., {Wu}, X.-B., {Fan}, X., {et~al.} 2018, \apj, 862, 109

\bibitem[{{Yang} {et~al.}(2019){Yang}, {Shen}, {Chen}, {Liu}, {Annis}, {Avila},
  {Bertin}, {Brooks}, {Buckley-Geer}, {Carnero Rosell}, {Carrasco Kind},
  {Carretero}, {da Costa}, {Desai}, {Diehl}, {Doel}, {Frieman},
  {Garcia-Bellido}, {Gaztanaga}, {Gerdes}, {Gruen}, {Gruendl}, {Gschwend},
  {Gutierrez}, {Hollowood}, {Honscheid}, {Hoyle}, {James}, {Krause}, {Kuehn},
  {Lidman}, {Lima}, {Maia}, {Marshall}, {Martini}, {Menanteau}, {Miquel},
  {Plazas Malagon}, {Sanchez}, {Scarpine}, {Schindler}, {Schubnell}, {Serrano},
  {Sevilla}, {Smith}, {Soares-Santos}, {Sobreira}, {Suchyta}, {Swanson},
  {Tarle}, {Vikram}, \& {Walker}}]{yang2019}
{Yang}, Q., {Shen}, Y., {Chen}, Y.-C., {et~al.} 2019, arXiv e-prints,
  arXiv:1904.10912

\bibitem[{{Zhang} {et~al.}(2018){Zhang}, {Du}, {Smith}, {Zhao}, {Hu}, {Xiao},
  {Li}, {Huang}, {Wang}, {Bai}, {Ho}, \& {Wang}}]{zhang2018}
{Zhang}, Z.-X., {Du}, P., {Smith}, P.~S., {et~al.} 2018, arXiv e-prints,
  arXiv:1811.03812

\end{thebibliography}



\end{document}